\documentclass[trackchanges, twocolumn]{aastex7}
\usepackage{ulem}

\usepackage{amsmath} 
\usepackage{comment}
\usepackage{appendix}
\usepackage[T1]{fontenc}
\usepackage[utf8]{inputenc}
\DeclareUnicodeCharacter{2013}{\textendash}   % –
\DeclareUnicodeCharacter{2014}{\textemdash}   % —
\DeclareUnicodeCharacter{2011}{-}             % non-breaking hyphen
\begin{document}

\title{Dynamical Masses and Radiative Transfer Modeling of HD\,698: a Be Binary in Evolutionary Transition}

\author[orcid=0000-0002-5328-8568,sname='Gabitova']{Ilfa A. Gabitova}
\affiliation{Faculty of Physics and Technology, Al-Farabi Kazakh National University, Al-Farabi Ave., 71, Almaty 050040, Kazakhstan}
\email{ilfa3110@gmail.com}  

\author[orcid=0000-0002-9369-574X, sname='Carciofi']{Alex C. Carciofi}
\affiliation{Instituto de Astronomia, Geof\'{i}sica e Ci\^encias Atmosf\'{e}ricas, Universidade de S\~ao Paulo, Rua do Mat\~ao 1226, Cidade Universit\'aria, 05508-900 S\~ao Paulo, SP, Brazil}
\email{carciofi@usp.br}

\author[orcid=0000-0001-5563-6629, sname='de Amorim']{Tajan H. de Amorim}
\affiliation{Instituto de Astronomia, Geof\'{i}sica e Ci\^encias Atmosf\'{e}ricas, Universidade de S\~ao Paulo, Rua do Mat\~ao 1226, Cidade Universit\'aria, 05508-900 S\~ao Paulo, SP, Brazil}
\email{tajan.amorim@usp.br}

\author[orcid=0000-0003-0696-2983, sname='Suffak']{Mark Suffak}
\affiliation{
Department of Physics and Astronomy, Western University, London, ON N6A 3K7, Canada}
\email{mark.suffak@ucalgary.ca}

\author[orcid=0000-0002-3833-1038, sname='Miroshnichenko']{Anatoly S. Miroshnichenko}
\affiliation{Department of Physics and Astronomy, University of North Carolina -- Greensboro, Greensboro, NC 27402, USA}
\affiliation{Fesenkov Astrophysical Institute, Observatory, 23, Almaty 050020, Kazakhstan}
\email{a\_mirosh@uncg.edu}

\author[orcid=0000-0003-2526-2683, sname='Zharikov']{Sergey V. Zharikov}
\affiliation{Instituto de Astronom\'ia, Universidad Nacional Aut\'onoma de M\'exico,  Apdo. Postal 106, Ensenada,  Baja California, M\'exico, 22860 }
\email{zhar@astro.unam.mx}

\author[orcid=0000-0002-2490-1562, sname='Rubio']{Amanda C. Rubio}
\affiliation{School of Physics and Astronomy, University of Leeds, Sir William Henry Bragg Building, Woodhouse
Ln., Leeds LS2 9JT, UK}
\email{a.c.rubio@leeds.ac.uk}

\author[orcid=0009-0001-7504-2438, sname='Danford']{Steve Danford}
\affiliation{Department of Physics and Astronomy, University of North Carolina -- Greensboro, Greensboro, NC 27402, USA}
\email{danford@uncg.edu}

\author[orcid=0000-0001-9788-7485, sname='Aarnio']{Alicia N. Aarnio}
\affiliation{Department of Physics and Astronomy, University of North Carolina -- Greensboro, Greensboro, NC 27402, USA}
\email{anaarnio@uncg.edu}

\author[sname='Prendergast']{Peter Prendergast}
\affiliation{Kernersville Observatory, Winston-Salem, NC 27285, USA}
\email{pgprendergast@yahoo.com}

\author[orcid=0000-0003-3096-759X,sname='Rudy']{Richard J. Rudy} 
\affiliation{Kookoosint Scientific, 1530 Calle Portada, Camarillo, CA 93010, USA}
\email{richardngc1068@gmail.com}

\author[orcid=0000-0002-2291-655X,sname='Puetter']{Richard C. Puetter}
\affiliation{Center for Astrophysics \& Space Science University of California, San Diego, 9500 Gilman Dr La Jolla, CA 92093, USA}
\email{rick.puetter@gmailcom}

\author[orcid=0000-0002-3389-1722,sname='Perry']{R. Brad Perry}
\affiliation{Alabaster Scientific, P.O. Box 120, Irvington, VA 22480, USA}
\email{rbradperry@icloud.com}

\author[orcid=0000-0001-9788-7485,sname='Agishev']{Aldiyar T. Agishev}
\affiliation{Faculty of Physics and Technology, Al-Farabi Kazakh National University, Al-Farabi Ave., 71, Almaty 050040, Kazakhstan}
\email{aldiyar.agishev@gmail.com} 

\author[orcid=0000-0002-7449-0108,sname='Vaidman']{Nadezhda L. Vaidman}
\affiliation{Faculty of Physics and Technology, Al-Farabi Kazakh National University, Al-Farabi Ave., 71, Almaty 050040, Kazakhstan}
\affiliation{Fesenkov Astrophysical Institute, Observatory, 23, Almaty 050020, Kazakhstan}
\email{nva1dmann@gmail.com} 

\author[orcid=0000-0001-5163-508X,sname='Khokhlov']{Serik A. Khokhlov}
\affiliation{Faculty of Physics and Technology, Al-Farabi Kazakh National University, Al-Farabi Ave., 71, Almaty 050040, Kazakhstan}
\affiliation{Fesenkov Astrophysical Institute, Observatory, 23, Almaty 050020, Kazakhstan}
\email{skhokh88@gmail.com}

\correspondingauthor{S.~A.~Khokhlov}
\email{skhokh88@gmail.com}

\begin{abstract}
We present a detailed analysis of the early post-mass-transfer binary system HD\,698 (V742~Cas), combining high-resolution optical spectroscopy, long-baseline interferometry, and radiative transfer modeling. Counter-phased RV curves reveal a circular orbit with a period of $55.927 \pm 0.001$~d and component masses of $M_{\rm Be} = 7.48 \pm 0.07~\rm M_{\odot}$ and $M_{\rm comp} = 1.23 \pm 0.02~\rm M_{\odot}$. The Be primary is traced via broad H$\alpha$ wings, while narrow metallic absorption lines originate from a slowly rotating companion. The angular separation measured via interferometry implies a dynamical distance of $888 \pm 5$~pc.

The SED is best reproduced with a color excess $E(B-V)=0.321 \pm 0.016$ due to interstellar reddening and a moderately dense viscous decretion disk with base density $\rho_0\simeq5\times10^{-12}\,\mathrm{g\,cm^{-3}}$ at $r=R_{\rm eq}$, declining radially as $\rho(r)\propto r^{-n}$ with $n=3.0$.
The companion is found to be a luminous and inflated star with $T_{\rm eff,comp} = 10.0^{+0.2}_{-0.1}$~kK, $R_{\rm comp} = 13.1^{+0.2}_{-0.2}~\rm R_\odot$, and $\log L/\rm L_{\odot} = 3.19$, contributing significantly to the flux ($L_{\mathrm{comp}}/L_{\mathrm{Be}} \sim 0.3$).

Spectral line mismatches provide further circumstantial evidence that the companion is hydrogen-poor, consistent with a stripped-envelope star enriched by CNO processing. HD\,698 thus belongs to the emerging class of Be + bloated O/B binaries, representing a short-lived, high-luminosity post-mass-transfer phase, when the stripped donor is still spectroscopically detectable before {reaching} the subdwarf phase.
\end{abstract}


\section{Introduction}

Classical Be stars are non-supergiant B-type stars that exhibit Balmer emission lines at some point in their lifetimes \citep{1982IAUS...98..261J}. Those emission lines, along with an infrared excess and an ultraviolet deficit, are typically attributed to a gaseous, nearly Keplerian circumstellar disk \citep{2013A&ARv..21...69R}. These stars are generally rapid rotators, whose equatorial velocities approach 70–80\% of the critical break-up speed. They exhibit spectroscopic and photometric variations associated with both stellar processes (such as non-radial pulsations and rotational modulation) and perturbations within their circumstellar environments.

Three principal scenarios have been proposed for the formation of Be star systems. The first suggests that Be stars are born as rapid rotators, retaining their angular momentum throughout the main sequence \citep{1997A&A...318..443Z}. The second scenario involves evolutionary spin-up during the main sequence. In this model, stars initially rotate at moderate speeds but experience internal angular momentum redistribution, leading to increased surface rotation rates as they evolve \citep{2008A&A...478..467E}. These two single-star evolutionary scenarios are supported by the presence of isolated Be stars \citep{2005ApJS..161..118M}.
The third is a binary evolution scenario, first introduced by \citet{1975BAICz..26...65K}. In this alternative theory, the Be phenomenon arises after mass and angular momentum transfer in an interacting binary, spinning up the initially less massive star to near-critical rotation \citep{1991A&A...241..419P, 2013ApJ...764..166D}. Both theoretical predictions and population studies have found support for this model. For example, \citet{2005ApJS..161..118M} concluded that $\sim$75\% of Be stars in young clusters may have undergone binary mass transfer.

If a classical Be star was spun up via binary interaction, theoretical models predict that after mass transfer, the donor star loses its hydrogen-rich envelope, becoming a stripped helium star \citep{2001A&A...369..939W, 2007A&A...467.1181D}. The outcome is a Be$+$He star binary, where the mass gainer displays rapid rotation and the mass donor is initially large and cold \citep[a bloated star,][]{2025A&A...694A.172R} that is progressively shrinking and heating up, evolving into an O or B class subdwarf (sdO/B) and eventually into a degenerate object. 

However, detecting companions in Be binaries remains difficult. The Be star’s broad lines, disk contamination, and brightness dominance often obscure the companions's spectroscopic or photometric signal. This selection bias may explain the lower observed binary fraction among fainter Be stars \citep{2011IAUS..272..304M}. In recent years the binary hypothesis has gained traction through infrared and interferometric studies. Disk truncation signatures in the spectral energy distributions (SEDs) of many Be stars are suggestive of tidal interactions with undetected companions \citep{2019ApJ...885..147K}. Spectroscopic surveys have confirmed that a substantial fraction of bright Be stars are binaries \citep{Wang_2021, 2022MNRAS.510.2286N}.

In recent years, a handful of newly identified Be binaries revealed the diversity of the post-mass-transfer phase. One such case is LB-1 (ALS~8775), originally reported as a Be star orbiting a $\sim$70~M$_\odot$ black hole \citep{2019ApJS..245...32L}. This interpretation was revised following spectral disentangling and UV analysis, which revealed the system to consist of a Be star and a $\sim$1~M$_\odot$ helium-rich pre-subdwarf companion \citep{2020A&A...633L...5I, 2020A&A...639L...6S}. The stripped star shows clear signatures of CNO-processed material, confirming its post-mass-transfer origin. A similar reevaluation occurred for HR~6819, first proposed to host a dormant black hole in a close orbit \citep{2020A&A...637L...3R}. Subsequent analyses and interferometric resolution showed that the system contains a classical Be star and a bloated companion, consistent with a recently stripped core of an intermediate-mass star \citep{2020A&A...641A..43B, 2021MNRAS.502.3436E}. In both cases, the low radial velocity (RV) of the Be star combined with its high rotational line broadening led to a misclassification of its physical properties and the mistaken inference of a black hole companion. One of the most recent such discoveries is HIP~15429. As reported by \citet{2025arXiv250406973M}, HIP~15429 is a newly identified Be+pre-He-star binary with an orbital period of 221~days. Its companion is a hot, compact remnant showing clear photospheric features, but still retains significant radius and luminosity. The system provides another example along the evolutionary sequence connecting Algol-type binaries, classical Be stars, and helium subdwarfs. 

LB-1, HR~6819 and HIP~15429 reveal a class of systems where the mass donor is caught in the transition toward the subdwarf sequence. In this configuration, the companion is still sufficiently luminous to be detected spectroscopically or interferometrically. This stage is expected to be short-lived (a few $\times 10^5$~yr) and may precede the hotter, more compact sdO/B phase seen in older Be binaries like $\phi$~Persei \citep{1998ApJ...493..440G}. Systems in this early phase offer rare observational access to both components immediately after mass transfer, before the companion evolves into a subdwarf.

HD\,698 (V742~Cas) has a long observational history as a peculiar and massive early-type binary. The system was first introduced by \citet{1932MNRAS..92..877P}, who derived its spectroscopic orbit using 40 spectra obtained over six years. They determined an orbital period of 55.9~days and identified the system as a binary composed of a B9 visible star and a B5-type companion. Based on interstellar Ca~\textsc{ii} absorption, Pearce estimated a distance of $\sim$1220~pc and an absolute magnitude of $M_V = -3.4$. His inferred minimum total mass of 158~M$_\odot$ was one of the largest ever proposed at the time. \citet{1948ApJ...108R.537S} revisited these findings, confirming the low orbital eccentricity ($e \approx 0$) and strong interstellar Ca~\textsc{ii} features, but questioned the extremely high luminosity and mass estimates. Using Stark-broadened hydrogen and helium line profiles, they revised the system’s absolute magnitude to a $M_V \sim -2$ to $-3$, implying that HD\,698 lies just above the main sequence. They argued that Pearce’s velocity measurements were likely affected by spectral line blending, and that the derived mass ratio ($q = 2.5$) and total mass were likely overestimated.

Subsequent studies attempted to clarify the nature of the system and its evolutionary state. \citet{1967LPlaS..77..219S} noted that the unseen companion appeared more massive than the visible component based on the mass function, suggesting HD\,698 as a prototype for binaries with hidden companions and hinting at the presence of an optically faint, massive object. He also detected H$\alpha$ emission that did not follow the orbital motion of the visible star, indicating the presence of a circumbinary or a disk-like envelope. Later, \citet{1978PASP...90..179H} conducted a high-resolution spectroscopic reinvestigation and again found no support for Pearce’s extreme mass claims. Instead, they estimated a more plausible component mass near 15~$M_\odot$ with a mass ratio closer to unity. The primary was classified as a B7~II--Ib giant with evidence for a stellar wind (mass-loss rate $\sim 10^{-6}~\rm M_\odot~ yr^{-1}$), while the companion remained spectroscopically undetected and at least 1.5~mag fainter than the primary. They also reported complex Balmer-line profiles, with H$\alpha$ appearing single-peaked and strong, while H$\beta$ showed a double-peaked shape, possibly tracing a rotating circumstellar structure. Next attempt to clarify the nature of the companion was made by \citet{1992ApJS...81..303S}, who analyzed International Ultraviolet Explorer's (IUE) spectra. Their work refined the orbital parameters and showed that the UV spectrum is dominated by the B5-type primary (T$_{\rm eff} \sim 15~000$~K), with no clear spectroscopic detection of a companion. Multiple sets of discrete absorption lines were identified. Most of them followed the orbital motion, others were attributed to non-thermal sources or to an expanding circumbinary shell. These findings supported the idea of circumstellar material and a complex radiation environment. The large mass function, combined with the persistent non-detection of the companion, led to speculation that HD\,698 might host a compact object such as a black hole \citep[see, e.g.,][]{2002ASPC..279..143C}, although no supporting X-ray emission or high-ionization features were ever found.

A breakthrough came with the recent work of \citet{2025A&A...694A.172R}, who combined long-baseline optical interferometry and high-resolution spectroscopy to resolve the nature of HD\,698’s elusive companion. Using the CHARA Array in the H band, they spatially resolved the binary and measured a projected separation of 0.663~mas, directly detecting the companion. Contrary to earlier expectations of a compact object, their analysis revealed that the companion is a low-mass, helium-rich pre-subdwarf—i.e., the stripped remnant of a once more massive star. The companion dominates the absorption-line spectrum in the optical, showing narrow metal lines consistent with slow rotation and photospheric abundances processed via CNO cycling. 

A tomographic study of HD\,698 was recently done by \citet{2025Galax..13...80G}, who analyzed H$\alpha$ and H$\beta$ emission profiles and mapped the structure of the circumstellar disk using Doppler tomography. Their analysis revealed violet-to-red (V/R) variations in both Balmer lines that are locked in phase with the orbital motion, suggesting a tidally perturbed, non-axisymmetric disk structure. The double-peaked profile is more prominent in H$\beta$, while H$\alpha$ appears more complex due to superimposed companion-related emission. Interestingly, in HD\,698 the V/R curve is phase-locked to the absorption-line RV curve in the same manner reported for a number of classical Be binaries \citep{2023Galax..11...83M}, despite these RVs trace the companion’s motion rather than that of the Be star.

In this paper, we investigated the physical properties of the HD\,698 system by combining new RV measurements from high-resolution optical spectra with recently published astrometric and interferometric data (Sect.~\ref{sec:data}). The orbital solution derived from these observations gives most precise to this date estimates for the system parameters, including the orbital period and component masses (Sect.~\ref{sec:orbit}). In addition, we performed radiative transfer modeling of the composite spectrum (Sect.~\ref{sec:modeling}), enabling us to constrain the parameters of the stripped companion. Finally, we place these results in the context of binary evolution and discuss their implications for the post-mass-transfer state of HD\,698 (Sect.~\ref{sec:discussion}) and summarize our findings in Sect.~\ref{sec:conclusions}.

%%%%%%%%%%%%%%%%%%%%%%%%%%%%%%%%%%%%%%%%%%
\section{Available data}\label{sec:data}

\subsection{Optical spectroscopy}

In this work, we built upon observations reported in \citet{2025Galax..13...80G} and used a total of 78 \'echelle spectra, including 71 spectra obtained between 2014 and 2025 at the Three College Observatory (TCO) and 7 spectra obtained at the Kernersville Observatory between 2016 and 2018. The observations cover a wavelength range of $\sim$ 3920~\AA\, to 7800~\AA, with a spectral resolving power of $R\approx12~000$. Both observatories were equipped with \'echelle spectrographs (Eshel from Shelyak Insruments\footnote{https://www.shelyak.com}) and ATIK--460EX detectors. Data reduction was performed %by one of us (A.M.) 
using standard procedures in IRAF, including bias subtraction, spectral order separation, and wavelength calibration with a ThAr comparison lamp. Flat-fielding was not performed because of a small variations of the pixel-to-pixel sensitivity ($\sim$1.5\%) of the detectors. More details on the data reduction are available in \citet{2023Galax..11....8M}.

\subsection{Spectral Energy Distribution}
\subsubsection{Photometry}
We compiled broad-band photometry for HD\,698 from multiple surveys using the VizieR catalog service \citep{Ochsenbein2000}. Two additional datapoints in the UV region were obtained from \citet{Jamar1976}. The full list of photometric data sources used in the study is provided in Table~\ref{tab:photometry}. 
In cases where multiple measurements were provided for one filter, the final value was considered to be the mean of the selected points with the standard deviation as the uncertainty.
Whenever just one data point was available for a filter and no uncertainty was provided, we adopted a conservative 10\% relative error. 

\begin{table*}[t]
\centering
\caption{Photometric data sources for HD\,698.}
\label{tab:photometry}
\begin{tabular}{lcc}
\hline
Survey & Bands & Reference \\
\hline
TD1 & FUV, NUV & \citet{Jamar1976} \\
Johnson & U, B, V, J, H, K, M, L & \citet{2006yCat.2122....0M} \\
Cousins & U, B, V, R, I & \citet{2006yCat.2122....0M} \\
SDSS & u, g, r, i & \citet{GaiaCollaboration2022} \\
HIP & $\rm H_p$, $\rm B_T$, $\rm V_T$ & \citet{ESA1997} \\
GAIA2 & $\rm G_{BP}$, G, $\rm G_{RP}$ & \citet{GaiaDR2} \\
GAIA3 & $\rm G_{BP}$, G, $\rm G_{RP}$ & \citet{GaiaDR3Spectra2023} \\
2MASS & J, H, $\rm K_s$ & \citet{Cutri2003} \\
WISE & W1, W2, W3 & \citet{Wright2010} \\
AKARI & S9W & \citet{Ishihara2010} \\
MSX & A & \citet{Egan2003} \\
PAN-STARRS/PS1 & g, r, i, z, y & \citet{2016arXiv161205560C} \\
Subaru/Suprime & IA598 & \citet{Miyazaki2012} \\
\hline
\end{tabular}
\end{table*}

\subsubsection{Gaia Spectrum and Astrometry}
For this study, we used astrometric and spectroscopic data from the Gaia Data Release \cite{GaiaCollaboration2022, GaiaDR3Spectra2023}. The spectroscopic data, specifically the BP/RP spectrum in the 330--1050~nm wavelength range, were retrieved from the Gaia DR3 archive\footnote{https://gea.esac.esa.int/archive/}. Resolving power $R\approx20-100$ depending on wavelength.

The astrometric solution, including parallax and proper motion, was obtained from the Gaia DR3 catalog. The parallax measurement of $1.4128 \pm 0.0358$~mas corresponds to a distance of $703 \pm 19$~pc \citep{2021yCat.1352....0B}. However, the renormalized unit weight error (RUWE) of 1.46 slightly exceeds the conventional threshold of 1.4 typically adopted to flag astrometric solutions that may be affected by unresolved source complexity or orbital motion \citep{2021A&A...649A...2L}.

\subsubsection{UV Spectroscopy}
We retrieved two ultraviolet spectra of HD\,698 from the IUE Newly Extracted Spectra (INES) database\footnote{https://sdc.cab.inta-csic.es/cgi-ines/IUEdbsMY}. Out of the nine spectra in the archive, we selected SWP~30287 (1151--1979~\AA ) and LWP~10100 (1851--3349~\AA ). Both were obtained using the large aperture. SWP~30287 is a spectrum with R$\approx$370 and the longest exposure time (7200~s) for this object. LWP~10100 has R$\approx$390 and low exposure time (25~s). We rejected the high-dispersion, long-exposure LWP~10101 spectrum from our study due to a noticeable shift of the 2175~\AA\ interstellar absorption feature. A detailed discussion of all IUE observations of HD\,698 can be found in \citet{1992ApJS...81..303S}.

\subsubsection{IR Spectroscopy}
We also incorporated a spectrum ($R\approx5~800$) covering a wavelengths range from 0.39 to 2.50~$\rm \mu m$. It was obtained with the Aerospace Near-Infrared Imaging Spectrograph \citep[NIRIS,][]{1999AJ....118..666R} on the 3--meter telescope at the Lick Observatory. This observation was conducted on 2018 October 20. A typical uncertainty of the flux measurements is $\sim$10\% across all wavelengths.

%%%%%%%%%%%%%%%%%%%%%%%%%%%%%%%%%%%%%%%%%%
\section{Orbital Solution}\label{sec:orbit}

\begin{figure*}[t]
\centering
\includegraphics[width=1\textwidth]{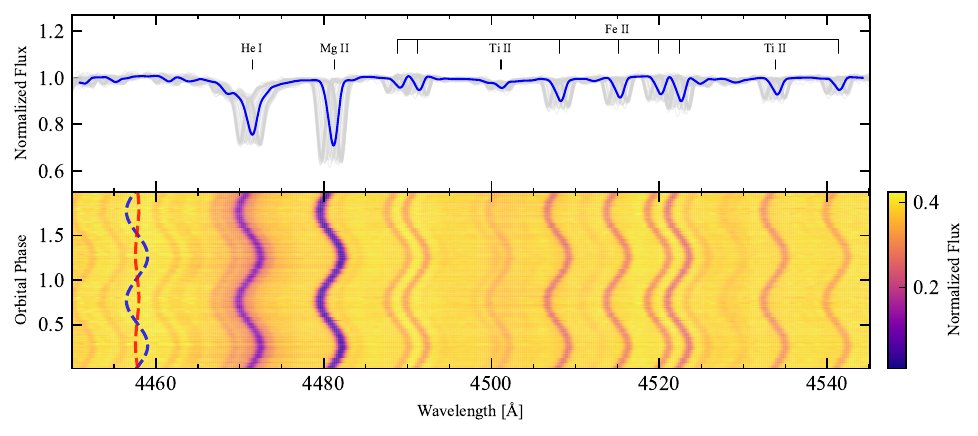}
\includegraphics[width=1\textwidth]{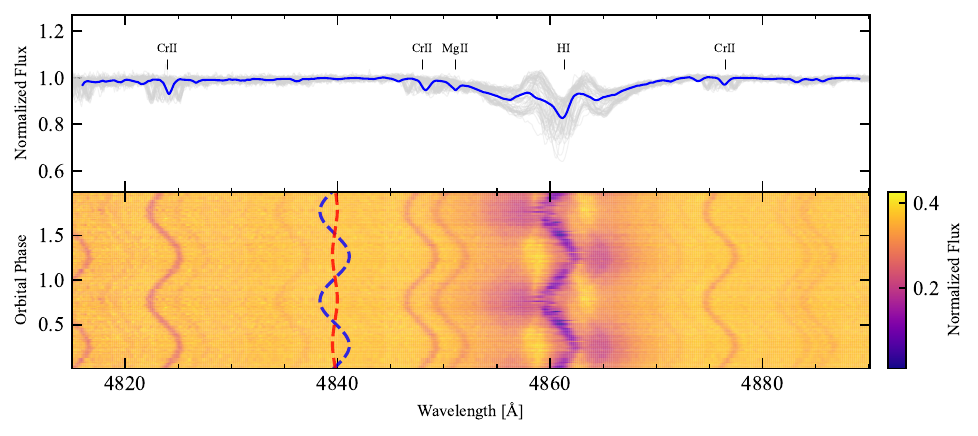}
\includegraphics[width=1\textwidth]{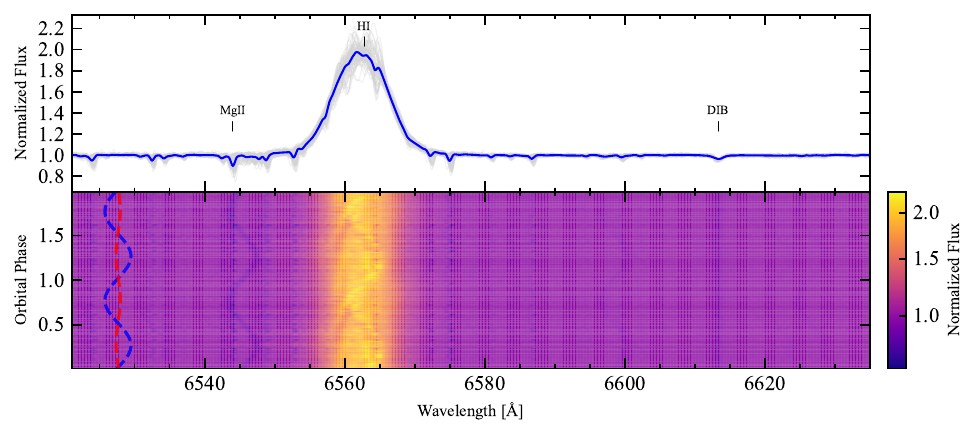}
\caption{Top panels: 78 spectra (grey) of the 4450--4545~\AA\ (upper plot), 4815--4890~\AA\ (middle plot), and 6520--6690~\AA\ regions (lower plot), overlaid with an averaged spectrum (blue). Bottom panels: phase-resolved spectra of the same regions. Blue dashed line demonstrates motion of the companion, red dashed line---motion of the Be star.}
\label{fig:dynspec}
\end{figure*}

The optical spectrum of HD\,698 is dominated by narrow absorption lines, which exhibit large-amplitude RV variations. Top panel of Figure~\ref{fig:dynspec} shows a phase-resolved spectrum of multiple metallic lines, demonstrating that all absorption features follow the same sinusoidal motion. No secondary set of photospheric lines is seen at any orbital phase. Historically, the visible RV curve was treated as arising from a Be star, and the companion was assumed to be spectroscopically invisible \citep[e.g.,][]{1932MNRAS..92..877P, 1948ApJ...108R.537S, 1978PASP...90..179H, 1991A&A...241..419P}.

Following the reinterpretation proposed by \citep{2025A&A...694A.172R}, we explore the alternative scenario in which the visible absorption lines arise from an unusually prominent low-mass companion, while the Be star's photospheric lines are obscured by the circumstellar disk and made shallow by rotational broadening. In this view, the high-amplitude RV curve corresponds to the low-mass companion. Supporting this interpretation, we measured a projected rotational velocity of $v \sin i = 28 \pm 5\rm~km~s^{-1}$ from 12 metallic lines in the 4481--4552~\AA\ region using the Fourier transform method \citep{2005A&A...440..305F}.

Adopting the interferometric inclination of $i = 119\fdg44 \pm 0\fdg27$ \citep{2025A&A...694A.172R}, this corresponds to an equatorial rotation speed of $v_{\rm rot,comp} = 32 \pm 6\rm~km~s^{-1}$. Such a slow rotation is inconsistent with a classical Be star, which typically rotates at several hundred km~s$^{-1}$, and would require an implausibly low inclination to reconcile with our observed value.

To derive the orbital parameters, we used 78 échelle spectra following \citet{2025Galax..13...80G} and measured RVs of the companion star via cross-correlation using the \texttt{xcsao} task in IRAF \citep{Kurtz_1998}, applied to the 4450--4545~\AA\ region containing 12 absorption lines (see the top panel of Fig. \ref{fig:dynspec}). To search for a spectroscopic signature of the Be star, we analyzed the wings of the H$\alpha$ emission line (Fig. \ref{fig:dynspec}), bottom panel) using a Gaussian fitting procedure. We fit a single Gaussian to the outer regions of the wings while excluding the central emission core. The fitting range was defined by a velocity mask, $v \leq v_\mathrm{min}$ or $v \geq v_\mathrm{max}$, to isolate the high-velocity regions presumed to trace the orbital motion of the inner disk. This method returns the centroid velocity of the wings for each epoch, interpreted as a proxy for the Be star’s RV. \footnote{We note that this method was cross-validated against several simpler estimators, including cutting wings at different amplitudes and using the mean velocity, and dislocating an exponential fit from one wing to another. These alternative methods reproduced our Gaussian-fit RVs with average difference from 0.47 to 1.78~km~s$^{-1}$.}

\begin{figure}[b]
\centering
\includegraphics[width=1\columnwidth]{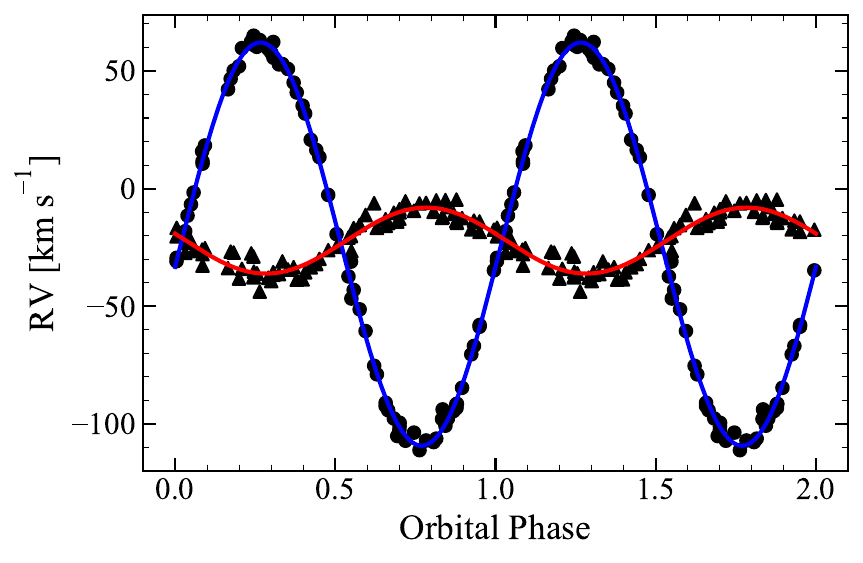}
\caption{Phase-folded RV curves: companion's RVs (circles), Be star's RVs (triangles), with fits overlaid (solid blue line for the companion and solid red line for the Be star). The average error of the companion's RV measurements is 1.67 km~s$^{-1}$, of the Be star's---2.07 km~s$^{-1}$.}
\label{fig:RV}
\end{figure}

RVs from both sources were then fitted using the \texttt{binarystarsolve} Python package \citep{2020arXiv201113914M} to determine the orbital solution.
The fitting resulted in a period of $P = 55.927 \pm 0.001$~d and time of companion's inferior conjunction $T_0 = 56050.946 \pm 0.154$~RJD\footnote{RJD = JD - 2400000}, matching the result of the previous study by \citet{2025A&A...694A.172R}. The RV semi-amplitude and systemic velocity are $K_{\rm comp} = 85.71 \pm 0.12$~km~s$^{-1}$ and $\gamma_{\rm comp} = -23.58 \pm 0.09$~km~s$^{-1}$, respectively. The eccentricity is $e = 0.005 \pm 0.002$. These parameters are in good agreement with previous studies, as can be seen in Table~\ref{tab:orbit}. Given the small eccentricity value and the consistency with the assumption of circularity in prior work \citep[e.g.,][]{2025A&A...694A.172R}, we adopt a circular orbit for all subsequent modeling and mass calculations. The Be star's RV curve is in anti-phase with that of the companion (See Fig. \ref{fig:RV}). The semi-amplitude $K_{\rm Be} = 14.04 \pm 0.11$~km~s$^{-1}$ and systemic velocity $\gamma_{\rm Be} = -22.12 \pm 0.08$~km~s$^{-1}$, in close agreement with the systemic velocity derived from companion’s motion.

\begin{table*}
\caption{Orbital parameters for HD\,698.}\label{params}
\begin{tabular}{lcccc} \hline
Parameter & [1] & [2] & [3] & This work \\
\hline
Number of observations & 11 & 31 & 29 & 78 \\
$P$ [d] & 55.9212 $\pm$ 0.0018 & 55.9233 $\pm$ 0.0009 & 55.9305 $\pm$ 0.0034 & 55.927 $\pm$ 0.001\\
$T_0$ [RJD] & 43315 $\pm$ 5 & 43316.585 $\pm$ 0.106 & 56050.939 $\pm$ 0.227 & 56050.946 $\pm$ 0.154\\
$e$ & 0.01 $\pm$ 0.01 & 0.005 $\pm$ 0.001 & 0.0 & 0 (fixed)\\
$K_{\rm Be}$ [km~s$^{-1}$] &  &  &  & 14.04 $\pm$ 0.11\\
$K_{\rm comp}$ [km~s$^{-1}$] & 84.2 $\pm$ 1.4 & 81.2 $\pm$ 0.96 & 87.06 $\pm$ 1.24 & 85.71 $\pm$ 0.12\\
$\gamma_{\rm Be}$ [km~s$^{-1}$] &  &  &  & $-$22.12 $\pm$ 0.08\\
$\gamma_{\rm comp}$ [km~s$^{-1}$] & $-25.6$ $\pm$ 0.9 & $-$27.1 $\pm$ 0.63 & $-$22.94 $\pm$ 0.87 & $-$23.58 $\pm$ 0.09\\
$q=M_{\rm comp}/M_{\rm Be}$ & 0.9--1.5 &  &  & 0.164 $\pm$ 0.001 \\
$M_{\rm Be}$ [M$_{\odot}]$ & 12.5--26.2 &  &  & 7.48 $\pm$ 0.07\\
$M_{\rm comp}$ [M$_{\odot}]$ & 11.3--39.2 &  &  & 1.23 $\pm$ 0.02\\
$i$ [$^\circ$] & 70--90 &  & 119.44 $\pm$ 0.27 & \\
$a$ [mas] &  &  & 0.663 $\pm$ 0.003 & \\
$d$ [pc] &  &  &  & 888 $\pm$ 5\\
\hline
\end{tabular} \label{tab:orbit}
\tablecomments{
        [1] - \cite{1978PASP...90..179H}. [2] - \cite{1992ApJS...81..303S}. [3] - \cite{2025A&A...694A.172R}.}
\end{table*}

The dynamical distance was obtained by comparing the projected orbital separation with the interferometric angular separation of $0.663 \pm 0.003$~mas \citep{2025A&A...694A.172R}, which gives a value of $888 \pm 5$~pc.
We argue that this estimate supersedes the Gaia parallax distance \citep[$703 \pm 19$~pc,][]{2021yCat.1352....0B}, given the marginal RUWE of 1.46 and the higher precision of our combined spectro-interferometric solution.
%%%%%%%%%%%%%%%%%%%%%%%%%%%%%%%%%%%%%%%%%%
\section{SED Modeling}\label{sec:modeling}

\subsection{Modeling Framework and Assumptions}
\begin{figure}[t]
\centering
\includegraphics[width=1\columnwidth]{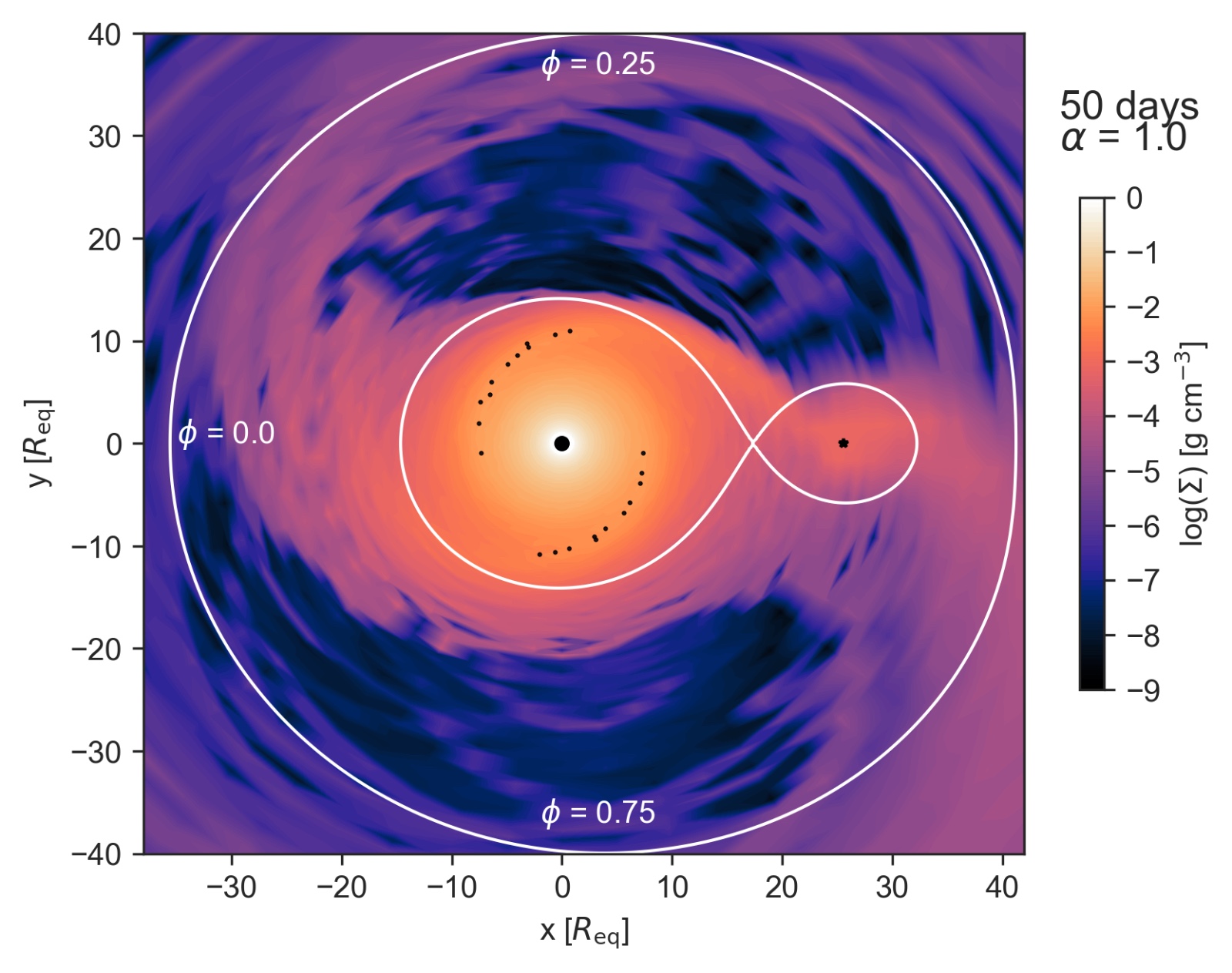}
\caption{Surface density map for a Be binary model with a 50-day orbital period, $q = 0.16$, and disk viscosity parameter $\alpha = 1.0$. The white lines trace the Roche equipotential contours, with the two large black circles representing the Be star (in x, y=0.0) and the companion (x=25.6~Req, y=0.0~Req). The small black dots trace the two spiral arms formed in the Be disk due to the presence of the companion.  More details about the simulation can be found in \citet{2025A&A...698A.309R}.}
\label{fig:schematic}
\end{figure}

To constrain the physical properties of HD\,698, we constructed a three-component spectral model that includes a Be star, its circumstellar disk, and a stripped companion. Figure~\ref{fig:schematic} illustrates the expected morphology of such a system, showing a model surface density map of a Be binary with similar orbital period and mass ratio. Given the complexity of the system and the degeneracies inherent in multi-parameter modeling, we adopted a hybrid approach: the Be star and the disk were constrained using physically motivated priors, while the companion's parameters were allowed to vary freely.

This strategy reflects the different evolutionary expectations for the two stars. The Be star is assumed to be near the terminal-age main sequence (TAMS), consistent with both the single-star spin-up and the binary mass-transfer formation channels. In either case, the Be phenomenon is expected to disappear as the star evolves off the main sequence and spins down. However, the duration and precise termination of the Be phase depend sensitively on the efficiency of internal angular-momentum transport and core–envelope coupling, which remain uncertain in models of massive stars \citep[e.g.,][]{2008A&A...478..467E, 2012A&A...537A.146E, 2012RvMP...84...25M}. Based on this, we adopted single-star rotating evolutionary tracks from the Geneva models \citep{2012A&A...537A.146E} to estimate the Be star’s temperature, radius, and luminosity given the mass obtained in Sect.~\ref{sec:orbit}. 
Specifically, we selected a solar-metallicity model with fractional main-sequence age $t/t_{\rm MS} = 0.95$ and rotational parameter $W = 0.8$, typical of classical Be stars \citep{2016A&A...595A.132Z}. The adopted parameters for the Be star are listed in the Table~\ref{tab:tabulated_params}.
In contrast, the companion is a stripped post-interaction object and is not expected to follow standard stellar evolution (see Sect.~\ref{sec:bpass}). We therefore treated its temperature and radius as free parameters, constrained by the observed SED.

For the circumstellar disk, a parametric version of the viscous decretion disk \citep[VDD,][]{10.1093/mnras/250.2.432} model was assumed.
Following a common approach in the literature \citep[e.g.,][]{2017MNRAS.464.3071V} the disk density follows a power-law in the equatorial plane and a Gaussian in the vertical direction:

\begin{equation}
\rho(r, z) = \rho_0 \left( \frac{R_{\rm eq}}{r} \right)^n \exp \left( - \frac{z^2}{2 H(r)^2} \right),
\end{equation}
where $r$ and $z$ are radial and vertical cylindrical coordinates respectively, in the stellar frame of reference, $\rho_0$ is the base density, and $H(r)$ is the vertical scale height, which increases with radius as:
\begin{equation}
H(r) = H_0 \left( \frac{r}{R_{\rm eq}} \right)^\beta,
\end{equation}
where $\beta$ is the flaring exponent. The base scale height $H_0$ is computed from the Be star’s temperature and mass, assuming an ideal gas. In addition, the velocity was assumed to be purely Keplerian with zero radial velocity and an azimuthal velocity given by \citet{2005ASPC..337...75B}.

Following \citet{2005ASPC..337...75B}, who derived approximate formulae for $\rho$ and $H$ under the assumption of an isothermal disk fed for an infinitely long time, we adopted a fixed flaring exponent $\beta = 1.5$ and a density fall-off index $n = 3.5$. In addition, we also considered a shallower slope, $n = 3.0$, to account for the mass accumulation effect expected in truncated disks of binaries \citep{2016MNRAS.461.2616P}.

For each value of the density slope (3.0  and 3.5), 
we considered seven values of the base density in the range $2.5\times10^{-12}-30\times10^{-12}\,\mathrm{g\,cm^{-3}}$. These values were guided by typical disk properties in Be star surveys \citep{2017MNRAS.464.3071V}.
This resulted in a set of 14 disk models for the fixed Be star mass. Finally, to explore distance-dependent effects, we computed models at discrete distances corresponding to the Gaia parallax and our dynamical estimate derived in Sect.~\ref{sec:orbit}.
Table~\ref{tab:tabulated_params} lists the parameters explored.

The Be star and disk spectra were precomputed using HDUST \citep{2006ApJ...639.1081C}, a three-dimensional Monte Carlo radiative transfer code that treats non-LTE conditions, rapid rotation, and circumstellar disk geometry. For each model, HDUST generates the emergent SED from far-UV to mid-IR, as well as H$\alpha$ and H$\beta$ line profiles. The stripped companion was modeled using interpolated  atmospheric models \citep{castelli_kurucz_2003_atlas9} over a regular grid in $T_\mathrm{eff}$ (3~500--50~000~K) and radius (1--30~R$_\odot$), with $\log g$ derived from the orbital mass and fitted radius. The model fluxes were reddened using the extinction law of \citet{1989ApJ...345..245C}, adopting a fixed $R_V = 3.1$ and allowing color excess $E(B-V)$ to vary between 0.01 and 0.95. The SED fitting served as the primary constraint (Sect.~\ref{sec:results}); line profiles were used as secondary consistency checks to evaluate physical plausibility (see below Sect.~\ref{sec:consistency}).

\subsection{MCMC Fitting and Parameter Exploration}
\label{sec:mcmc_fitting}

To determine the physical parameters of HD\,698, we performed a grid-based Markov Chain Monte Carlo (MCMC) fit to the observed SED. The dataset combines broad-band photometry and low-resolution spectra spanning ultraviolet, optical, and infrared wavelengths (see Sect.~\ref{sec:data}).

The total observed flux ($F^{\mathrm{total}}_{\lambda}$) at wavelength $\lambda$ is modeled as the sum of the intrinsic fluxes from the Be star with its circumstellar disk ($F^{\mathrm{Be+disk}}_\lambda$) and from the companion star ($F^{\mathrm{comp}}_\lambda$) as
\begin{equation}
\begin{split}
F^{\mathrm{total}}_{\lambda} = [ F^{\mathrm{Be+disk}}_\lambda(\rho_0, n, d) + \\
+ F^{\mathrm{comp}}_\lambda (T_{\mathrm{eff,comp}}, R_{\mathrm{comp}}, d)] \cdot 10^{-0.4~ A_\lambda(E(B{-}V), R_V)}\,.
\end{split}
\end{equation}
%\begin{equation}
%F_\lambda = [F^{1}_\lambda(\rho_0,n,d) + F^{2}_\lambda(T_{2}^{eff},R_2)]10^{-0.4A_\lambda}
%\end{equation}

\noindent Both contributions are scaled to the system's distance ($d$) and then attenuated by interstellar extinction, represented by the multiplicative factor $10^{-0.4A_\lambda(E(B-V),R_V)}$. The companion's effective temperature, $T_\mathrm{eff,~comp}$ and radius, $R_\mathrm{comp}$ were treated as free parameters, along with color excess $E(B-V)$. Distance $d$ and disk parameters, the slope $n$ and base density $\rho_0$, composed an external grid, as shown in Table~\ref{tab:tabulated_params}. For each of the 28 resulting model–distance combinations, we ran an independent MCMC chain using the \texttt{emcee} package \citep{ForemanMackey2013} to explore the posterior distributions of the free parameters.

\begin{table*}[t]
\centering
\caption{Explored grid of tabulated parameters used in SED modeling.}
\label{tab:tabulated_params}
\begin{tabular}{ll}
\hline
\textbf{Parameter} & \textbf{Explored Values} \\
\hline
\textbf{Be star (adopted parameters):} \\
\hline
Mass $M_{\rm  Be}~[\rm M_\odot]$ & 7.5 \\
Radius $R_{\rm Be}~[\rm R_\odot]$ & 6.3 \\
Effective Temperature $T_{\rm{eff,Be}}$ [kK] & 20.1 \\
Surface gravity $\log g$ [cgs] & 3.7 \\
Luminosity $\log L~[{\rm L_\odot}]$ & 3.8 \\
\hline
\textbf{Circumstellar Disk:} \\
\hline
Density slope $n$ & 3.0;\ 3.5 \\
Base density $\rho_0$ [$10^{-12}\,\mathrm{g\,cm^{-3}}$] & 2.5;\ 5.0;\ 7.5;\ 10;\ 15;\ 20;\ 30 \\
\hline
\textbf{System-Wide Parameter:} \\
\hline
Distance $d$ [pc] & 703;\ 888\\
\hline
\end{tabular}
\end{table*}

A $\chi^2$ statistic was used to assess goodness of the final SEDs:
\begin{equation}
\chi^2_{\mathrm{k}} = \sum_{i=1}^{N} \left( \frac{ F_{\lambda,i}^{\mathrm{obs}} - F_{\lambda,i}^{\mathrm{model}}}{\sigma_i} \right)^2,
\end{equation}
where both $i$ indexes over a total number $N$ of either photometric points or binned spectral fluxes; $F_{\lambda,i}^{\mathrm{obs}}$ represents observed flux and $F_{\lambda,i}^{\mathrm{model}}$---modeled one, $\sigma_i$ is the observational uncertainty.
To ensure that no wavelength range dominated the fit, we grouped the residuals into five spectral intervals (UV, Balmer jump, optical, NIR, and IR) and normalized their contributions to give each region equal weight in the total $\chi^2$, that was then constructed as the mean of the region-wise contributions:
\begin{equation}
\chi^2 = \frac{1}{N_k} \sum_{k=1}^{N_k} \chi^2_k\,,
\end{equation}
with $N_k = 5$.

\subsection{Best-Fit Model and Results}
\label{sec:results}

Following the MCMC fitting procedure, we recomputed the final SED of the stripped companion using HDUST, adopting the most probable parameters ($T_{\mathrm{eff}}$ and $R_{\rm comp}$) from the posterior distributions. This step was necessary to ensure full consistency with the radiative transfer treatment applied to the Be star and disk components, particularly for the generation of high-resolution line profiles and scattered flux contributions. The companion's synthetic SED, produced in HDUST, replaced the Kurucz-based grid spectra used during MCMC sampling in the final best-fit model.

\begin{figure*}
\centering
\includegraphics[width=1.0\textwidth]{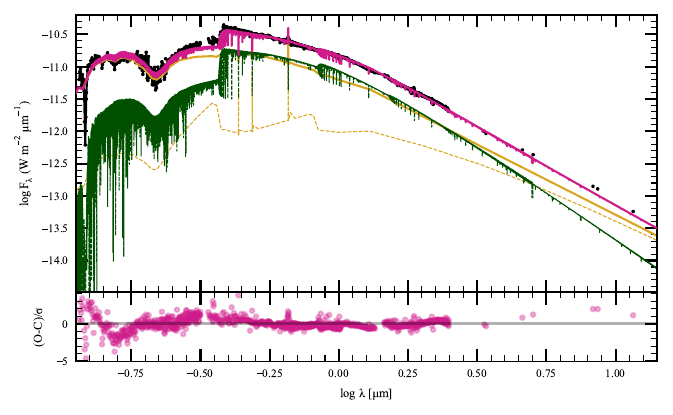}
\caption{Top panel: The total modeled SED (purple solid line) is the sum of the Be star and its disk (yellow solid line) and companion's (green dashed line) contributions, both generated with HDUST. Observational data is shown in black. Yellow dashed line represents flux contribution from the disk. Bottom panel: residuals distribution across wavelengths.
}
\label{fig:sedbest}
\end{figure*}

Figure~\ref{fig:sedbest} shows the synthetic SED corresponding to the highest posterior probability model from our MCMC analysis across all sampled combinations of distance, disk density structure, extinction, and companion properties. In the top panel, the total SED (purple solid line) is decomposed into contributions from the Be star and its circumstellar disk (yellow solid line) and the stripped companion (green solid line), with all components modeled using HDUST. 
The disk contribution to the SED is shown as a yellow dashed line.
The observed data are plotted in black. The bottom panel shows the residuals across wavelength. The residuals are generally contained within $\pm1\sigma$ through most of the optical and NIR regions. Around the Balmer jump and in the MIR residuals are skewed positive at the level of $\sim1$–$2\sigma$. The former might be a sign of a mismatch in the modeled hydrogen abundance, and the latter might be hinting that a disk contribution is slightly too faint in this regime.

The associated parameters for this model are summarized in Table~\ref{tab:bestfit}. Notably, the derived color excess, $E(B-V) = 0.322^{+0.002}_{-0.002}$, is in close agreement with the value from the Bayestar19 3D dust map \citep{green2019_bayestar19} at the dynamical distance of 888~pc ($E(B-V) = 0.309$), and the value obtained by \citet{1992ApJS...81..303S} from the IUE spectra ($E(B-V) = 0.36$).
The companion star is significantly inflated, with a radius of $R_{\rm comp} = 13.1^{+0.2}_{-0.2}$~R$_\odot$, but relatively cool, with an effective temperature of $T_{\rm eff,comp} = 10.0^{+0.2}_{-0.1}$~kK. Despite this, its luminosity is comparable to that of the Be primary ($L_{\mathrm{comp}}/L_{\mathrm{Be}} \sim 0.3$). The significant temperature difference between the two stars explains why the Be star dominates the ultraviolet part of the spectrum, while both stars contribute comparably in the optical. In the infrared, the Be star again dominates, due to emission from its circumstellar disk, characterized by a base density of $\rho_0 = 5.0\times10^{-12}\,\mathrm{g\,cm^{-3}}$ and a density slope of $n = 3.0$.

\begin{figure*}
\centering
\includegraphics[width=0.32\textwidth]{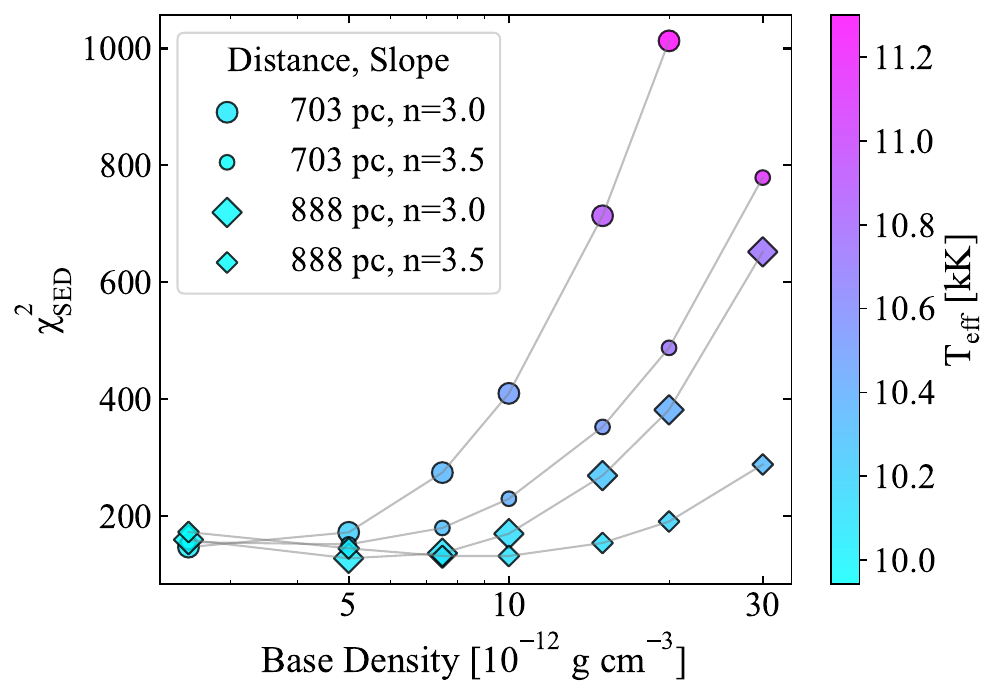}
\includegraphics[width=0.32\textwidth]{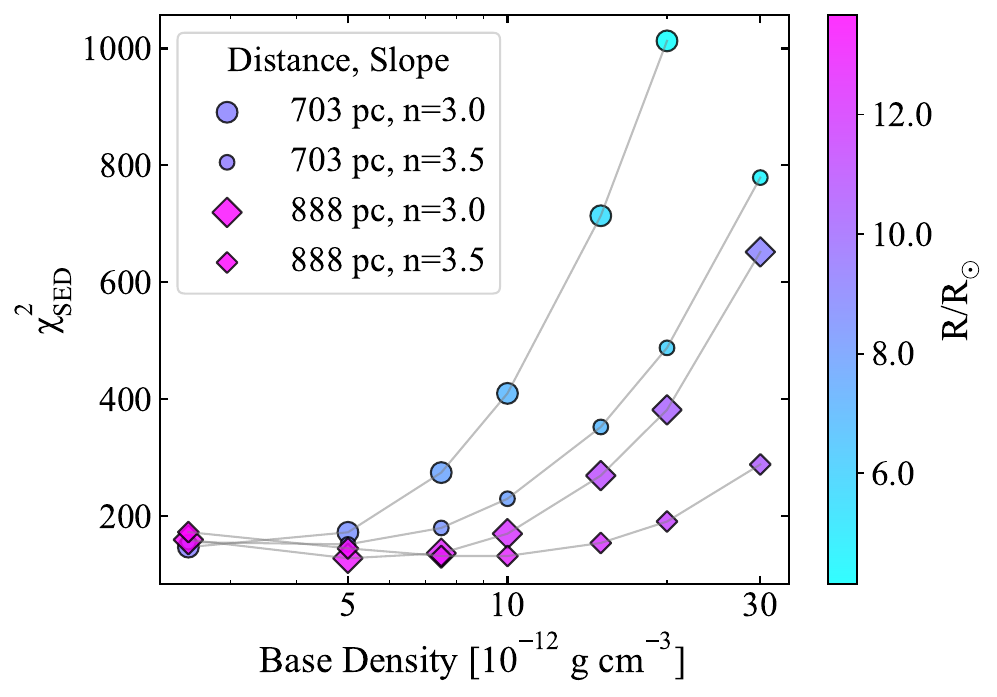}
\includegraphics[width=0.32\textwidth]{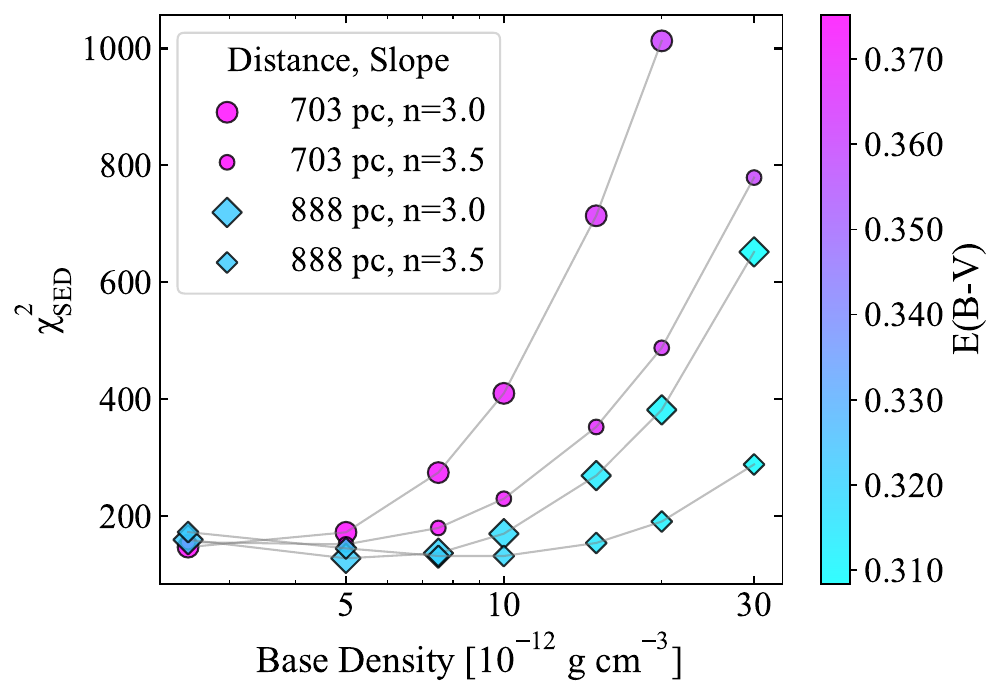}
\caption{Distribution of SED $\chi^2$ values as a function of base density for HDUST models at two distances (703~pc and 888~pc) and two density slope parameters (n=3.0 and n=3.5). Each panel shows the same data colored by different free parameters: companion's effective temperature (left panel), companion's radius (central panel), and interstellar extinction $E(B-V)$ (right panel). Diamonds represent models at 703 pc, circles represent models at 888 pc. Larger symbols correspond to n=3.0, smaller symbols to n=3.5. Gray lines connect models with identical distance and slope parameters to highlight density-dependent trends. Lower $\chi^2$ values indicate better fits to the observed SED.}
\label{fig:chi}
\end{figure*}

Figure~\ref{fig:chi} shows that the model quality, as measured by total SED $\chi^2$, shows dependencies on free parameters, disk parameters, and the distance. The fit favors intermediate base densities in the range $\rho_0 = 5\times10^{-12}$ to $10\times10^{-12}\,\mathrm{g\,cm^{-3}}$. As can be seen in Fig.~\ref{fig:sedbest}, the IR is where disk's contribution is most notable. Lower values of disk's base density underpredict the IR excess, while higher densities begin to overproduce it. Regarding the companion’s physical parameters, higher $T_{\mathrm{eff}}$ values produce excess UV flux and degrade the fit, particularly below 3000~\AA, where the Be star dominates the spectrum.
There is a preference for models with larger companion radii, but this trend must be interpreted in conjunction with the effective temperature. Bigger radii and lower temperatures provide better agreement in the near-IR and optical, where the flux contribution from the companion becomes non-negligible. Conversely, models with small radii and high temperatures tend to contribute insufficiently to the flux in these regions.

\begin{table}[b]
\centering
\caption{Most Probable Solution from SED Radiative Transfer Modeling.}
\label{tab:bestfit}
\begin{tabular}{lcc}
\hline
\textbf{Parameter} & \textbf{Type} & \textbf{Value}\\
\hline
\textbf{Companion:} \\
\hline
$R_{\rm comp} [\rm R_\odot]$ & free & $13.1^{+0.2}_{-0.2}$ \\
$T_{\rm eff,comp}$ [kK] & free & $10.0^{+0.2}_{-0.1}$ \\
\hline
\textbf{Circumstellar Disk:} \\
\hline
$n$ & discrete & 3.0 \\
$\rho_0$ $[10^{-12} \rm ~{g~cm}^{-3}]$ & discrete & 5.0 \\
\hline
\textbf{System-Wide Parameters:} \\
\hline
$d$ [pc] & discrete & 888 \\
$E(B-V)$ & free & $0.322^{+0.002}_{-0.002}$ \\
\hline
\end{tabular}
\end{table}

These correlations indicate that the companion is bloated, relatively cool, and luminous. Although individual parameter uncertainties are small in the MCMC posteriors, these reflect the local shape of the likelihood and do not account for broader systematic effects. The strength of our modeling lies in excluding implausible solutions. The MCMC posteriors (Appendix~\ref{sec:appendix}; Figs.~\ref{fig:A1}–\ref{fig:A7}) constrict the preferred solutions to $T_{\rm eff}\approx 10.0$--$11.2$~kK and $R\approx 4.4$--$13.7~R_\odot$. An extremely cool/compact case (Fig.~\ref{fig:A7}, $T_{\rm eff}\sim 3.6$~kK, $R\sim 3.1~R_\odot$) remains a single outlier with the highest residuals and was omitted from the $\chi^2$ summary plots (Fig.~\ref{fig:chi}) to preserve the scale. If the dynamical distance alone is adopted ($d=888$~pc), the posteriors tighten further to $T_{\rm eff}\!\approx\!10.0$–$10.7$~kK and $R\!\approx\!9.1$–$13.7$~R$_\odot$. Thus, while absolute values remain subject to model assumptions (composition, disk symmetry), our analysis tightly constrains the companion's effective temperature and, while radius values vary in a bigger range, compact solutions are ruled out.

\subsubsection{Consistency of Line Diagnostics with SED Models}
\label{sec:consistency}

In this subsection, we discuss the line profile predictions of the SED-based modeling outlined above. Even though no attempt has been made to incorporate the line profiles as modeling constraints---for the reasons discussed at the end of this subsection--- this comparison is nevertheless useful.

\begin{figure}[t]
\includegraphics[width=0.49\columnwidth,trim=5 3 5 3, clip]{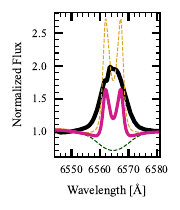}
\includegraphics[width=0.49\columnwidth,trim=5 3 5 3, clip]{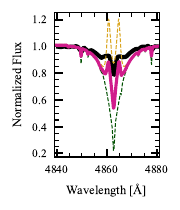}
\includegraphics[width=0.49\columnwidth,trim=5 3 5 3, clip]{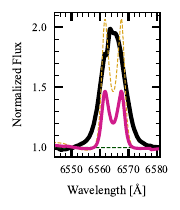}
\includegraphics[width=0.49\columnwidth,trim=5 3 5 3, clip]{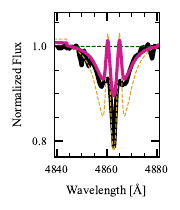}
\caption{
Top panels: comparison of HDUST-generated most-probable solution model (purple solid line) with the average observed H$\alpha$ (left panel) and H$\beta$ (right panel) line profiles (black solid line). The model includes contributions from the Be star and its disk (yellow dashed line) and the companion (green dashed line). Bottom panels: same, but companion's flux assumed to be flat. For each panel, we report the EWs for comparison: 
top–left (H$\alpha$): EW$_{\mathrm{observed}}= -9.47$\,\AA, EW$_{\mathrm{model}}= -2.61$\,\AA; 
top–right (H$\beta$): EW$_{\mathrm{observed}}= 1.10$\,\AA, EW$_{\mathrm{model}}= 2.39$\,\AA; 
bottom–left (H$\alpha$): EW$_{\mathrm{observed}}= -9.47$\,\AA, EW$_{\mathrm{model}}=-4.51$\,\AA; 
bottom–right (H$\beta$): EW$_{\mathrm{observed}}= 1.10$\,\AA, EW$_{\mathrm{model}}= 0.31$\,\AA.}
\label{fig:balmer_diagnostics}
\end{figure}

Although our SED modeling tightly constrains global system parameters, the synthetic spectra fail to reproduce the observed spectral lines. 
Figure~\ref{fig:balmer_diagnostics} shows the model and observed Balmer line profiles with and without the companion's contribution. Models across the parameter space systematically predict lower EWs with both H$\alpha$ and H$\beta$.
Removing the companion's absorption component---to mimic a hydrogen free atmosphere---improves the H$\alpha$ profile but the EW is still lower than the observed H$\alpha$'s value, while H$\beta$'s EW becomes overestimated.
This suggests that the companion may be hydrogen-poor, consistent with expectations for a stripped post-interaction object. Testing this hypothesis will require spectroscopic modeling with non-solar abundances. 
Variations in disk density, companion's effective temperature, or radius do not resolve the discrepancy within the limits of solar-abundance models.

\begin{figure}[t]
\includegraphics[width=1.0\columnwidth]{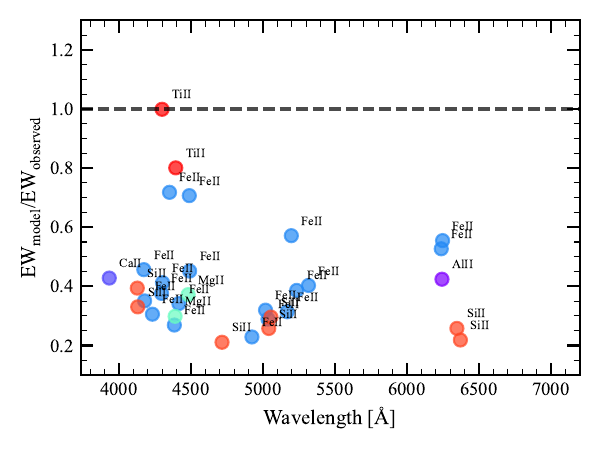}
\caption{EW comparison of HDUST-generated absorption lines with the observed spectrum. Black dashed line indicates perfect fit.}
\label{fig:abs_diagnostics}
\end{figure}

A complementary comparison was performed for selected photospheric absorption lines of the companion. Observed equivalent widths (EWs) were measured in one high-SNR echelle spectrum, corrected for the systemic velocity. 
Figure~\ref{fig:abs_diagnostics} compares measured EWs to those predicted by the best-fit SED model. In all cases, the model predicts weaker absorption lines, not within observational uncertainties. Assuming smaller companion radii or higher temperatures results in only minor improvements, and the mismatch persists across the parameter grid. Models computed at Gaia’s parallax distance slightly improve the Balmer line fits while slightly worsening the absorption line fits; however, these changes do not resolve the discrepancies.

The poor performance of line diagnostics highlights two key limitations of our current models. First, the assumption of solar composition likely fails for the stripped companion, which may exhibit strong CNO-cycle enrichment and an altered opacity structure. Second, the circumstellar disk model assumes axial symmetry and no external irradiation, which likely underestimates Balmer emission. It was shown by \citet{2025Galax..13...80G} that the emission lines in HD\,698 have a complex structure, significantly influenced by the companion. These effects should be addressed in future modeling using non-solar abundance atmospheres and more complex hydrodynamical disk structures.

\section{Discussion} \label{sec:discussion}

\subsection{Nature of the System}

HD\,698 appears to be a member of the emerging class of Be + bloated binaries. This rare configuration has only been recognized in the past few years \citep[e.g., LB-1 (LS~V+22~25), HR~6819, and HIP~15429,][]{2020A&A...639L...6S, 2021MNRAS.502.3436E, 2025arXiv250406973M}. These systems are now understood as post-interaction Be binaries with recently stripped stars rather than compact objects as companions. As summarized in Table~\ref{tab:comparison}, HD\,698 shares the defining traits with LB-1, HR~6819, and HIP~15429: a rapidly rotating Be primary and a low-mass luminous companion.

\begin{table*}
\caption{Comparison of HD\,698 with known Be+bloated O/B binaries.}
\label{tab:comparison}
\begin{tabular}{lcccc}
\hline
\textbf{Parameter} & \textbf{HD\,698} & \textbf{LB-1} [1] & \textbf{HR~6819} [2]& \textbf{HIP~15429} [3]\\
\hline
\multicolumn{5}{l}{\textbf{Be Star}} \\
\hline
Spectral Type &  & B3 Ve & & B3V \\
$M$ [${\rm M_\odot}$] & 7.48 $\pm$ 0.07 & 5 & $6.7^{+1.9}_{-1.5}$ & $\geq$ 7.0\\
$T_{\rm eff}$ [kK] & 20.1 & 18 & $18^{+2}_{-3}$ & $17^{+2}_{-1}$ \\
$R$ [$\rm R_\odot$] & 6.3 & 3.7 & $4.7^{+2.7}_{-1.5}$ & $6.0^{+1.6}_{-1.3}$ \\
$\log g$ [cgs] & 3.7 & 4.0 & $3.75^{+0.5}_{-0.25}$ & 4.0 $\pm$ 0.5 \\
$\log L~[{\rm L_\odot}]$ & 3.8 & 3.1 & $3.35^{+0.47}_{-0.44}$ & $3.36^{+0.20}_{-0.19}$ \\
$v \sin i$ [km~s$^{-1}$] & 332 & 300 $\pm$ 50 & 180 $\pm$ 20 & 270 $\pm$ 70\\
\hline
\multicolumn{5}{l}{\textbf{Bloated Companion}} \\
\hline
Spectral Type &  &  &  & B5Ib \\
$M$ [{$\rm M_\odot$}] & 1.23 $\pm$ 0.02 & 1.1 & $0.47^{+0.28}_{-0.22}$ & \\
$T_{\rm eff}$ [kK] & $10.0^{+0.2}_{-0.1}$ & 12.7 & 16 $\pm$ 1 & 13.5 $\pm$ 0.5 \\
$R$ [$\rm R_\odot$] & $13.1^{+0.2}_{-0.2}$ & 5.4 & $4.7^{+2.9}_{-1.9}$ & $9.0^{+2.1}_{-1.7}$ \\
$\log g$ [cgs] & 2.29 $\pm$ 0.02 & 3.0 & 2.75 $\pm$ 0.35 & 2.25 $\pm$ 0.25\\
$\log L~[{\rm L_\odot}]$ &  $3.19^{+0.05}_{-0.03}$ & 2.8 & $3.11^{+0.42}_{-0.46}$ & $3.43^{+0.20}_{-0.23}$ \\
$v \sin i$ [km~s$^{-1}$] & 28 $\pm$ 5 & 7 $\pm$ 2 & $<$20 & $\leq$30 \\
\hline
\multicolumn{5}{l}{\textbf{System-wide}} \\
\hline
$P_{\rm orb}$ [days] & 55.927 $\pm$ 0.001 & 78.7999 $\pm$ 0.0097 & 40.3 $\pm$ 0.3 & 221 $\pm$ 1\\
$e$ & 0 & 0 & $<$0.037 & 0.52 $\pm$ 0.03\\
$q = M_{\rm comp}/M_{\rm Be}$ & 0.164 $\pm$ 0.001 & 0.21 $\pm$ 0.02 & 0.071 $\pm$ 0.032 & 0.066 $\pm$ 0.053\\
$d$ [pc] & 888 $\pm$ 5 &  &  & 1730 $\pm$ 260 \\
\hline
\end{tabular}
\tablecomments{
        1--LB-1 \citep{2020A&A...639L...6S},
        2--HR~6819 \citep{2021MNRAS.502.3436E},
        3--HIP~15429 \citep{2025arXiv250406973M}. We homogenized literature values to a uniform convention (in particular, mass ratio $q$ is defined as $M_{\rm comp}/M_{\rm Be}$ and luminosity is given as $\log L~[{\rm L_\odot}]$). Small symmetric uncertainties were propagated with first-order differentials. In non-linear or asymmetric cases we transformed the published bounds directly. For masses and mass ratio, we favored dynamical constraints over spectroscopic or evolutionary ones.
        }
\end{table*}

\begin{figure}[t]
\includegraphics[width=1.0\columnwidth]{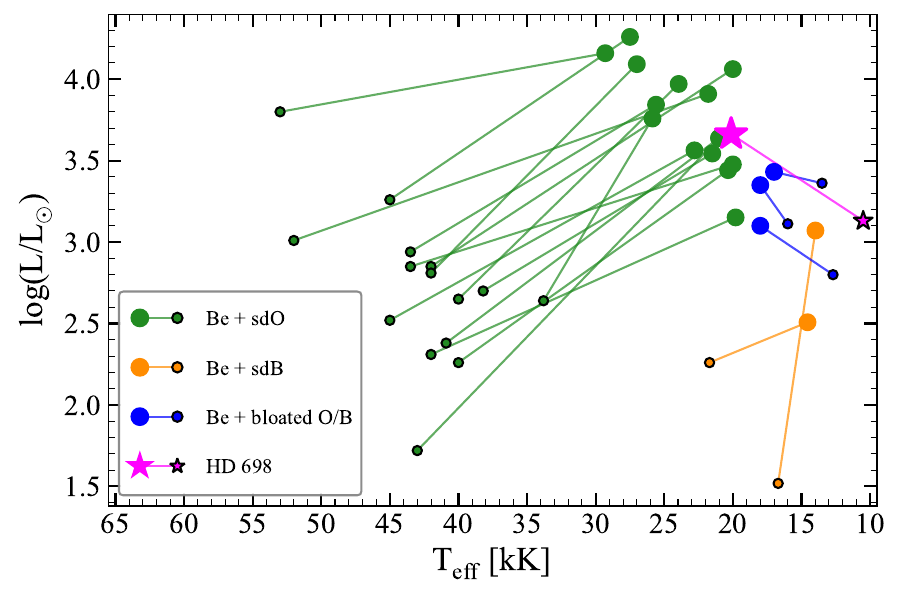}
\caption{Comparison of HD\,698's components (magenta stars) on HR diagram with known Be+sdO binaries (green circles) \citep{Gies_1998, refId0, Wang_2017, Wang_2021, Klement_2024, Peters_2008, Peters_2013, Peters_2016, Chojnowski_2018}, Be+sdB binaries (orange circles) \citep{Klement_2022, 2025arXiv250720053D} and three example Be+bloated O/B binaries from Table~\ref{tab:comparison} (blue circles).
}
\label{fig:hrd}
\end{figure}

Figure~\ref{fig:hrd} illustrates that the known Be + bloated-companion systems 
occupy a region at intermediate effective temperatures ($T_{\rm eff}\sim10$--$20$~kK) and high luminosities for their masses. These companions are likely expanded stripped cores of previously more massive stars. Classical Be + sdO systems \citep[e.g., $\phi$~Persei][]{Gies_1998, refId0} host compact, hot ($T_{\rm eff}>30$~kK) remnants, located far to the left on the HR diagram.
HD\,698 and the other examples listed above lie well apart, forming a distinct branch of Be binaries. 

Beyond individual Be+bloated O/B binaries, population studies show that binary interaction can produce a large number of stripped helium stars with present-day masses of roughly 2--8~$\rm M_\odot$ and $T_{\rm eff}$ of 50--100\,kK \citep{2023Sci...382.1287D}. These stars are bridging subdwarfs and Wolf–Rayet stars and may be revealed by their UV excess. At somewhat earlier evolutionary stages, Be/Oe+partially stripped binaries with overluminous, nitrogen-enhanced donors with masses of 4--8~$\rm M_\odot$ and $T_{\rm eff}$ of 21--40\,kK have now been identified in the Magellanic Clouds \citep{2024A&A...692A..90R}. \textcolor{blue}{The companions in Table\,\ref{tab:comparison}, however, are cooler and less massive than the systems explored in those studies. At the same time, these companions, particularly in HD\,698 and LB-1 systems are more massive than expected from canonical sdO/B companions in Be binaries, which typically cluster around 0.5~$\rm M_\odot$. However, a few well-studied cases demonstrate that significantly higher-mass remnants can occur. FY~CMa, for example, hosts a 1.1--1.5~$\rm M_\odot$ subdwarf \citep{Peters_2008}, and V2119~Cyg's and 60~Cyg’s subdwarfs are reported to have masses of 1.35~$\rm M_\odot$ and 1.45~$\rm M_\odot$ respectively \citep{Klement_2024}. The $1.23 \pm 0.02~\rm M_\odot$ companion of HD\,698 therefore lies comfortably within the observed range of companions' masses in Be binaries, regardless of its precise evolutionary stage or whether further mass-loss episodes will occur.}

\subsection{Evolutionary status of the system}
\label{sec:bpass}

To test our interpretation and gain insight into the possible progenitor characteristics of the system, as well as its future evolution, we used the Binary Population and Spectral Synthesis code \citep[BPASS,][]{eld17} and the Hoki package \citep{ste20} to search for evolutionary tracks of systems that would undergo mass transfer and result in B stars on the MS and lower-mass stripped stars. We searched the BPASS binary models at solar metallicity (Z = 0.02), and filtered for those that have undergone mass transfer and result in a mass-gainer star with a mass between 7.4 to 7.6 $\rm M_\odot$, and a low mass companion (0.2 to 1.25 $\rm M_\odot$), the ranges of which are consistent with our determined masses in Table~\ref{tab:orbit}. 
We also excluded systems in which the final orbital period was shorter than the initial one, as this implies highly non-conservative mass transfer. A total of 12 BPASS models satisfied these selection criteria.

\begin{table}
    \caption{BPASS model parameters.}
    \label{tab:BPASS_params}
    \centering
    \begin{tabular}{lcc}
    \hline
    \textbf{Parameter} & \textbf{Model 1} & \textbf{Model 2} \\
    \hline
    Initial Donor Mass [$\rm M_\odot$] & 5 & 4.5 \\
    Initial Gainer Mass [$\rm M_\odot$] & 4.5 & 3.6 \\
    Initial Orbital Period [days] & 3.98 & 2.51 \\
    Donor Mass at HD\,698 [$\rm M_\odot$] & 4.93 & 0.68 \\
    Donor Mass at sdO/B Stage [$\rm M_\odot$] & 0.92 & 0.68 \\
    \hline
    \end{tabular}
\end{table}

\begin{figure*}
    \includegraphics[width=0.33\textwidth]{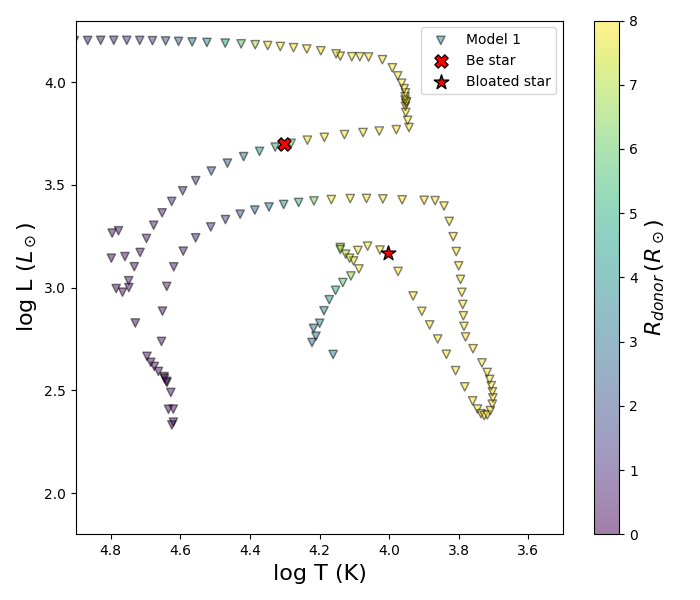} 
    \includegraphics[width=0.33\textwidth]{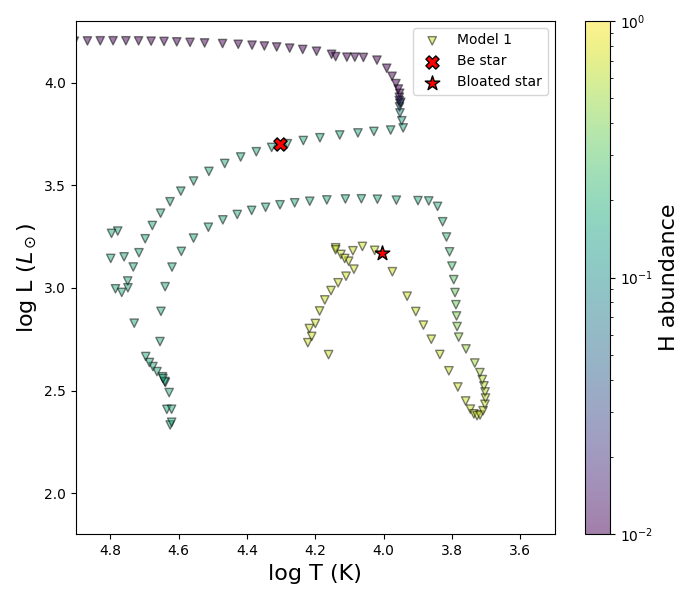} 
    \includegraphics[width=0.33\textwidth]{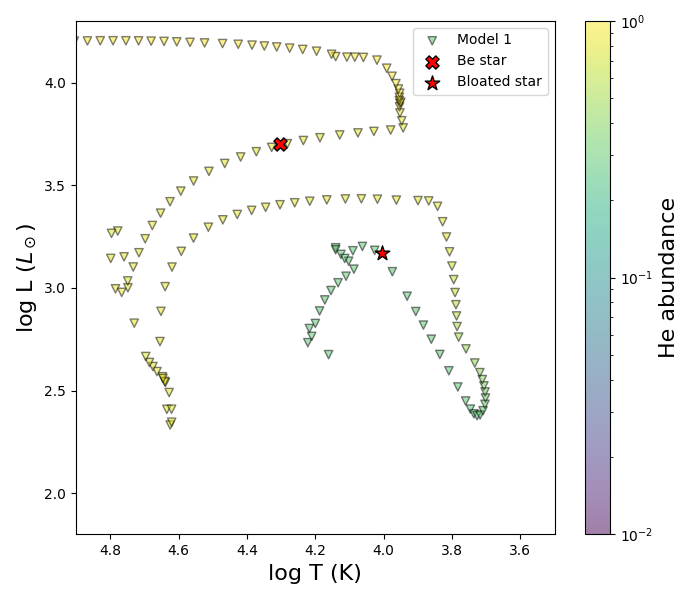} 
    \includegraphics[width=0.33\textwidth]{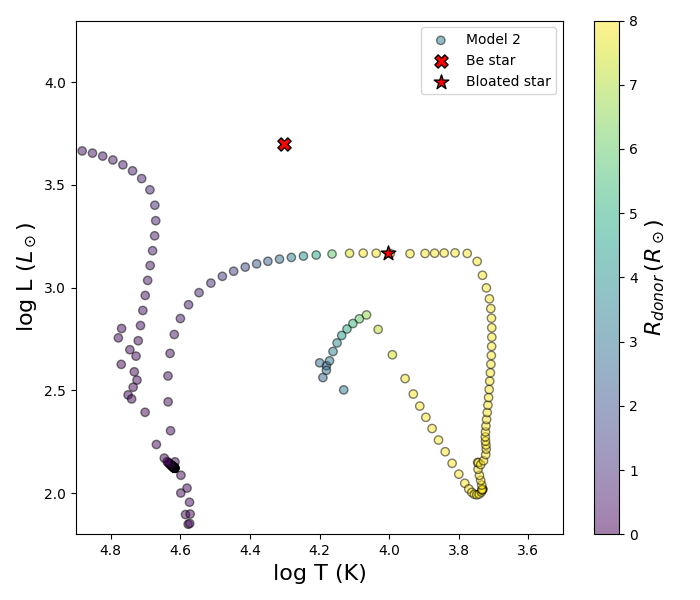} 
    \includegraphics[width=0.33\textwidth]{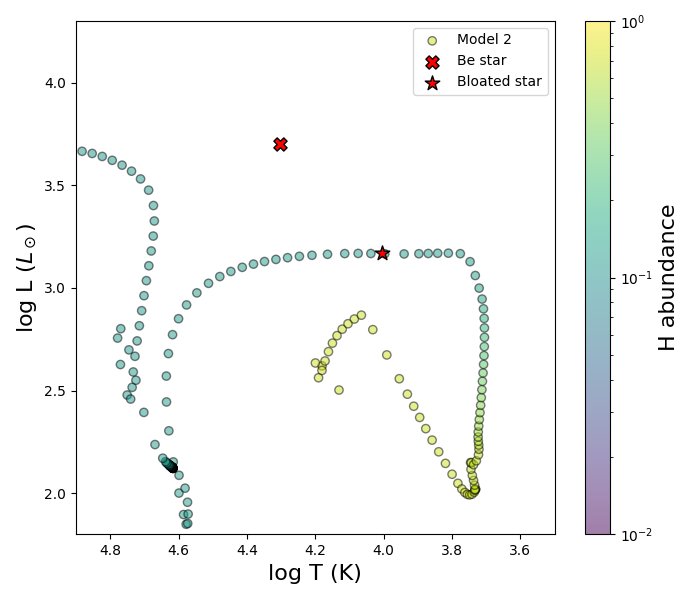} 
    \includegraphics[width=0.33\textwidth]{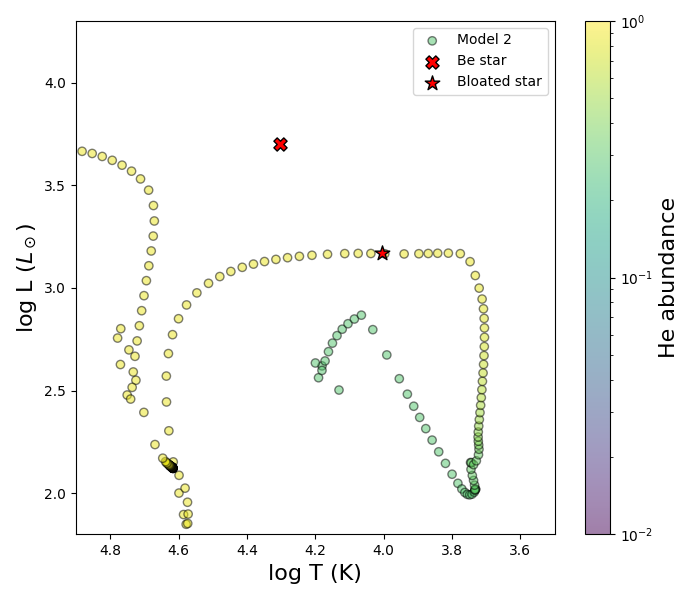} 
    \caption{HR diagrams with evolutionary tracks of the mass-donor from two BPASS binary evolution models. The location of HR\,698 is given by the red `x' and star symbols for the Be and bloated component, respectively. The two BPASS models of Table~\ref{tab:BPASS_params} are shown in the top and bottom rows, respectively. Left to right, the tracks are colored by the mass-donor's radius, Hydrogen abundance, and Helium abundance. %\commentMS{Not sure if both tracks on a single plot is too busy/hard to interpret. Could make more panels and just have single tracks if that is preferred.}
    }
    \label{fig:BPASS_models}
\end{figure*}
The evolutionary track of the mass-donor in two of the filtered BPASS models overlapped directly with the location of the bloated star of HR~698 on the HR diagram. These two tracks are plotted in Figure~\ref{fig:BPASS_models} along with the location of both components of HR~698. Parameters describing both models are listed in Table~\ref{tab:BPASS_params}.

For the two models shown, the donor star first evolves off the main sequence, leaving the terminal-age main sequence (TAMS) at a moderate luminosity. As core hydrogen is exhausted, the star expands and cools, moving rightward across the HR diagram. This expansion triggers mass transfer to the companion, during which the donor becomes increasingly distended.

After this bloated phase, the star evolves leftward at a nearly constant luminosity. Its effective temperature rises significantly as the outer envelope is lost, and the donor’s radius shrinks. At this stage, the surface composition is already modified by prior mass loss, and the star transitions toward a hot subdwarf configuration.
Finally, the track settles at a high temperature and a relatively low luminosity, marking the stripped-envelope remnant. The star at this stage is helium-rich at the surface and much more compact than during its bloated phase. 

The tracks in the first column of Fig.~\ref{fig:BPASS_models} are colored by the radius of the donor star. The companion star to HD\,698 (red star in the figure) lies in the first bloated portion of both tracks, consistent with our previous determination; however, it intersects the tracks at distinctly different places: 
for model 1 (triangles, top-left) it does so
near the start of the bloated phase and for model 2 (circles, bottom-left), near the end. 
In the first case, the bloated star would have just recently left the main sequence and would be ``moving right'' across the HR diagram. The intersection happens $\approx 2 ~ \rm Myr$ after the end of the main sequence.
For model 2, it would be ``moving left'' towards the sdO/sdB stage of evolution. At that point, $\approx 19 ~ \rm Myr$ has passed since the end of the main sequence.
%\addACC{This, by itself, favors Model 2 as the more plausible scenario, since in Model 1 the secondary’s mass, at the phase corresponding to the position of the bloated star in the HR diagram, is too large to be explained.}
To distinguish between these two scenarios, note in Table~\ref{tab:BPASS_params} that the mass of the donor is still large when Model 1 intersects the location of the bloated star, while in Model 2 mass transfer has been nearly completed by the time of intersection. This points towards Model 2 being more accurate to the evolutionary status of HR\,698. %; however, surface abundance measurements will be a more telling constraint.

%So then how would we be able to differentiate between the two tracks and determine whether the bloated star is moving right or left across the HR diagram? This question can be answered through the middle and bottom panels of Figure~\ref{fig:BPASS_models}, where the evolutionary tracks are colored by hydrogen and helium abundance, respectively. The difference is then clear, in that shortly after the main-sequence, the bloated companion is still mostly hydrogen, while when the bloated star is moving left towards the subdwarf stage, helium is much higher than hydrogen. Therefore, surface abundance measurements of the bloated star would allow researchers to refine the evolutionary stage of the companion to HR~698.

To further distinguish between the two evolutionary tracks and determine whether the bloated star is moving rightward or leftward across the HR diagram, we examine the second and third columns of Fig.~\ref{fig:BPASS_models}, where the tracks are color-coded by hydrogen and helium abundance, respectively. The distinction is clear: shortly after the main sequence, the bloated companion is hydrogen-rich, whereas during its leftward evolution toward the subdwarf stage, helium dominates. Surface abundance measurements of the bloated star can therefore constrain the evolutionary stage of the companion to HR\,698. 

Recent work by \citet{2025A&A...693A..10S} has clarified an important distinction between stellar expansion (or contraction) and inflation (or deflation). Expansion results from a thermal imbalance that develops after core-hydrogen exhaustion, when the nuclear luminosity and the actual luminosity diverge and the envelope readjusts on a thermal timescale. By contrast, inflation refers to the development of tenuous, radiation-dominated layers near the local Eddington limit, which can occur even in thermally balanced models and is characterized by low-density envelopes supported primarily by radiative pressure. The removal of such inflated layers, while the star remains in thermal equilibrium, should properly be described as deflation rather than contraction.

Although a detailed internal-structure analysis is beyond the scope of the present paper, the evolutionary interpretation of HD\,698 may hinge on this distinction. Provided that the stripped companion is indeed evolving leftward in the Hertzsprung–Russell diagram, i.e., toward higher effective temperatures and smaller radii, its present evolution could involve either deflation of a previously inflated, radiation-supported envelope, or a more conventional contraction following the cessation of shell-hydrogen burning and restoration of thermal balance. Discriminating between these two scenarios will require dedicated stellar-structure modeling that follows the envelope response to both thermal and radiative-pressure instabilities in the post-mass-transfer phase.
%%%%%%%%%%%%%%%%%%%%%%%%%%%%%%%%%%%%%%%%%%

\section{Conclusions}\label{sec:conclusions}

The nature of HD\,698, as consisting of a classical Be star and a low-mass bloated companion, is confirmed through the analysis of RV data and spectral features. This RV curve was historically attributed to the Be star, assuming an unseen companion. However, we adopt the scenario proposed by \citet{2025A&A...694A.172R} in which the absorption lines originate from an unusually luminous companion, while the Be star’s lines are obscured by its circumstellar disk and rotationally broadened. The Be star is traced through the wings of H$\alpha$ emission. The resulting orbital solution gives a mass ratio of $q = 0.164 \pm 0.001$, with component masses of $M_{\rm Be} = 7.48 \pm 0.07~\rm M_\odot$ and $M_{\rm comp} = 1.23 \pm 0.02~\rm M_\odot$.

Using our spectroscopic orbit and the interferometric angular separation given by \citet{2025A&A...694A.172R}, we derive a dynamical distance of $888 \pm 5$~pc. This distance is preferred over the Gaia parallax by our radiative transfer modeling as well.

The Be star is modeled as a near–TAMS object with $M_{\rm Be} = 7.5~\rm M_\odot$, $T_{\rm eff,Be} = 20.1$~kK, and $R_{\rm Be} = 6.3~\rm R_\odot$. Its disk is best reproduced by a density profile with $\rho_0 = 5.0 \times 10^{-12}~\mathrm{g~cm^{-3}}$ and $n = 3.0$, truncated at the Roche lobe. Radiative transfer modeling of the spectral energy distribution also confirms that the companion is an inflated, luminous, low-mass object. We find $T_{\rm eff,comp} = 10.0^{+0.2}_{-0.1}$~kK and $R_{\rm comp} = 13.1^{+0.2}_{-0.2}~\rm R_\odot$, with $\log g = 2.29 \pm 0.02$. These properties are incompatible with both main-sequence stars and compact subdwarfs, and instead indicate a post-mass-transfer transient evolutionary stage.

While the SED fit accurately reproduces the system's broadband fluxes, synthetic line profiles systematically underestimate both Balmer and metallic line strengths. This likely reflects the limitations of solar-composition models and the use of axisymmetric, non-irradiated disk structures. These discrepancies motivate follow-up modeling with non-solar abundances and time-dependent hydrodynamics.

Evolutionary tracks from BPASS indicate that the companion is a recently stripped donor star in the bloated phase, transitioning towards the subdwarf stage. HD\,698 thus joins the small but growing list of post-mass-transfer Be binaries representing a transient evolutionary stage. These systems offer rare constraints on the immediate aftermath of mass transfer and the pathway to classical Be + sdOB binaries, such as $\phi$~Persei.

The novelty of this study lies not only in the precision of the orbital solution, but also in the construction of the first self-consistent Be + bloated star model. This establishes a benchmark for future investigations of post-interaction Be binaries, where the secondary can no longer be treated as an elusive or negligible contributor to the system’s energy budget.

Looking ahead, an important step will be a detailed spectroscopic analysis of the secondary. Such a study, however, cannot be undertaken in isolation: as demonstrated here, the secondary must be modeled jointly with the primary and its circumstellar disk to obtain reliable physical parameters. This combined approach will be essential for disentangling the spectra of both components, constraining surface abundances and evolutionary models, and ultimately advancing our understanding of the post-mass-transfer fate of massive binaries.

\begin{acknowledgments}
This research was funded by the Science Committee of the Ministry of Science and Higher Education of the Republic of Kazakhstan (grant No. AP23484898).

This research has made use of the SIMBAD database, operated at CDS, Strasbourg, France; SAO/NASA ADS, and Gaia data products.

A. C. C. acknowledges support from CNPq (grant 314545/2023-9) and FAPESP (grants 2018/04055-8 and 2019/13354-1). 

T. H. A. acknowledges support from FAPESP (grant 2021/01891-2).

S. V. Zh. acknowledges the DGAPA-PAPIIT grants IN119323, IN105826.

This work made use of the computing facilities of the Centro de Processamento de Dados do IAG/USP (CPD-IAG), whose purchase was made possible by the Brazilian agency FAPESP (grants 2019/25950-8, 2017/24954-4 and 2009/54006-4).

We thank Jorge Sahade for the IUE observations.

Finally, we thank the anonymous referee for their comments and suggestions, which improved the manuscript.
\end{acknowledgments}

\clearpage
\appendix

\setcounter{figure}{0}
\renewcommand{\thefigure}{A\arabic{figure}}
\makeatletter
\renewcommand{\theHfigure}{A\arabic{figure}}
\makeatother

\section{MCMC Corner Plots}\label{sec:appendix}
\begin{figure*}
\centering
\includegraphics[width=0.45\textwidth]{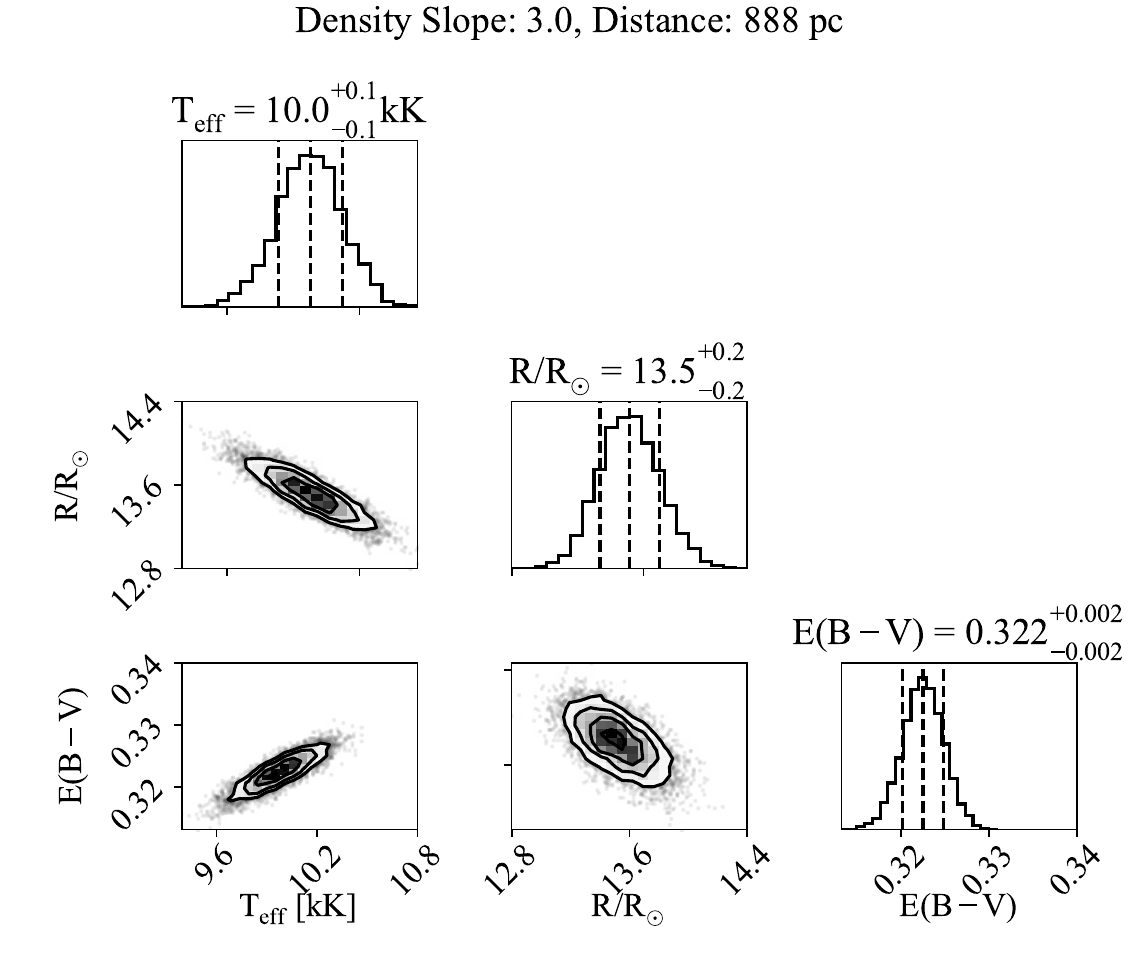}
\includegraphics[width=0.45\textwidth]{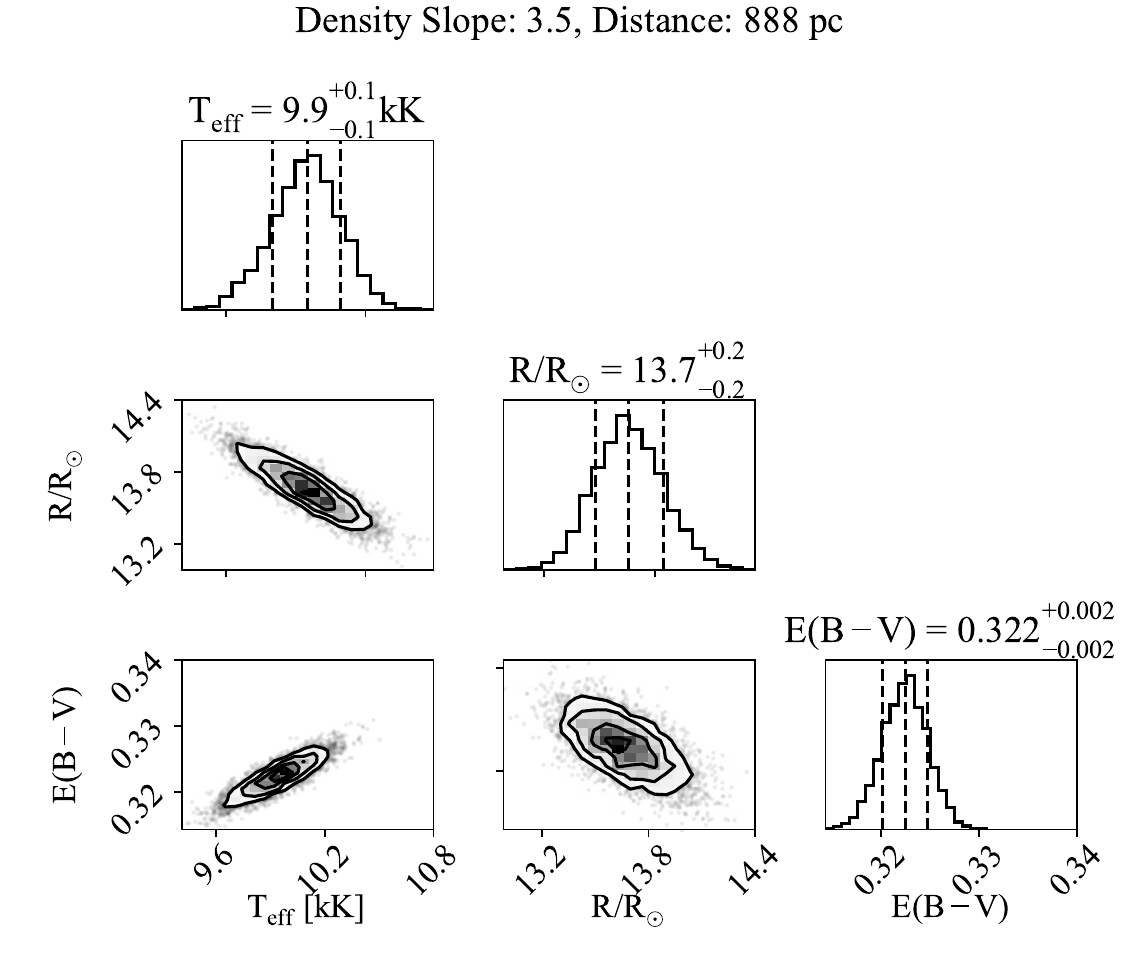}
\includegraphics[width=0.45\textwidth]{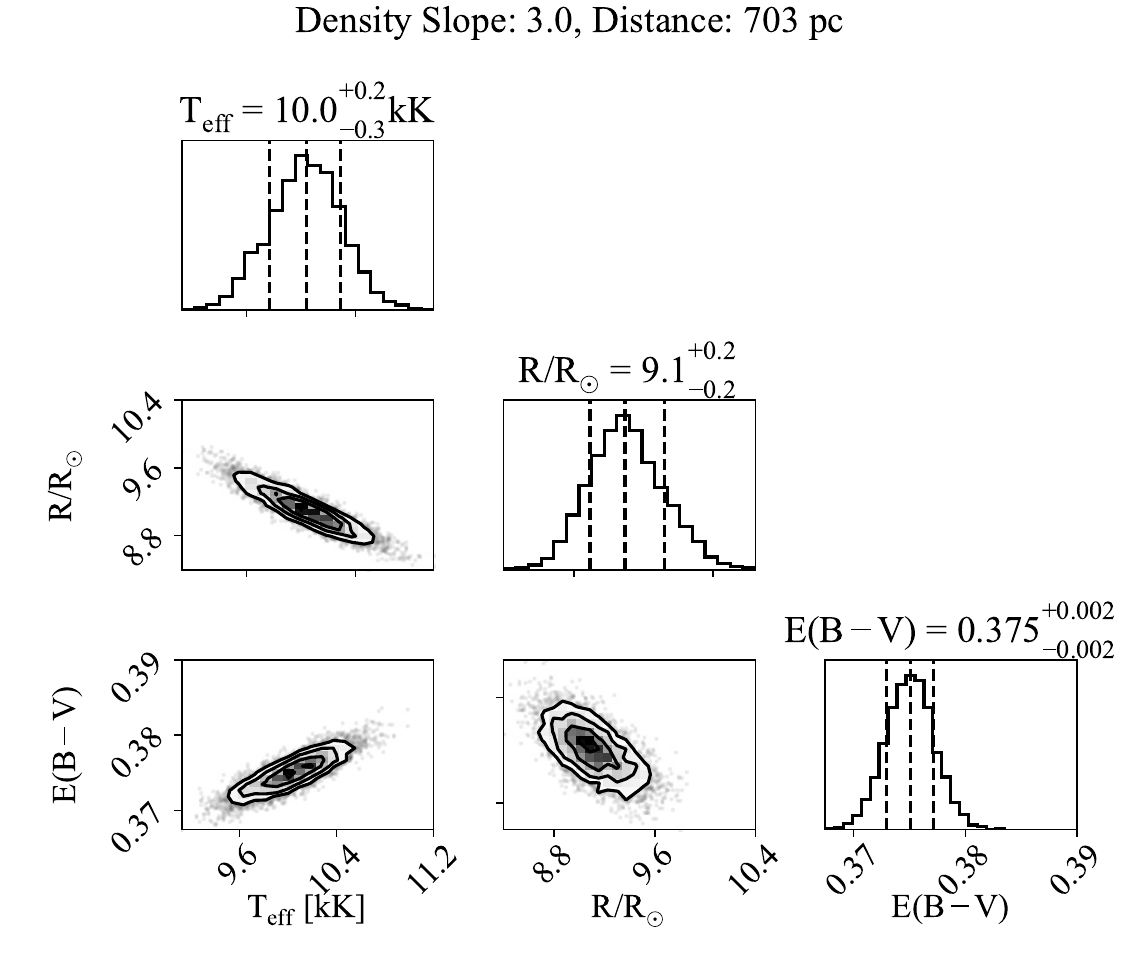}
\includegraphics[width=0.45\textwidth]{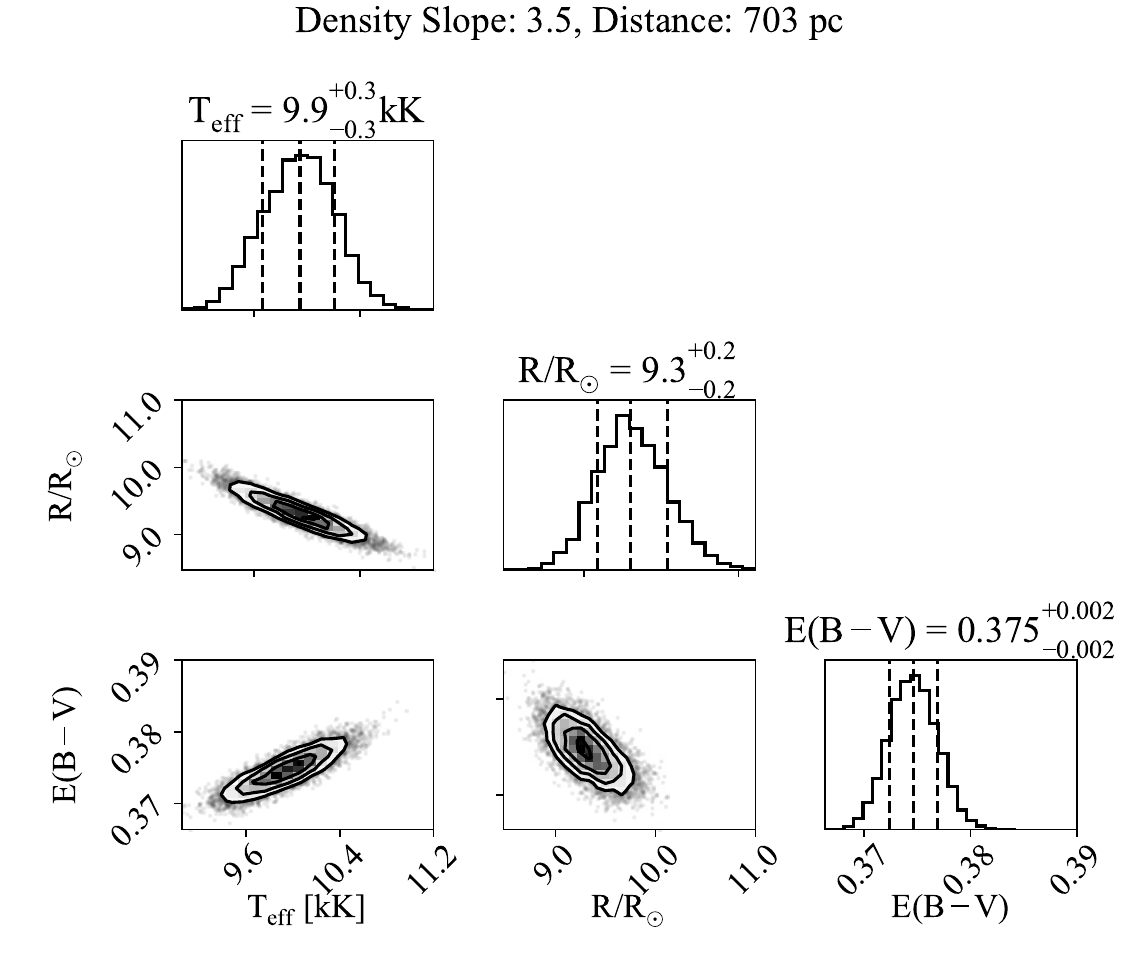}
\caption{Corner plots of the MCMC posteriors for the model with disk base density $\rho_0 = 2.5\times 10^{-12}\,{\rm g\,cm^{-3}}$.}
\label{fig:A1}
\end{figure*}

\begin{figure*}
\centering
\includegraphics[width=0.45\textwidth]{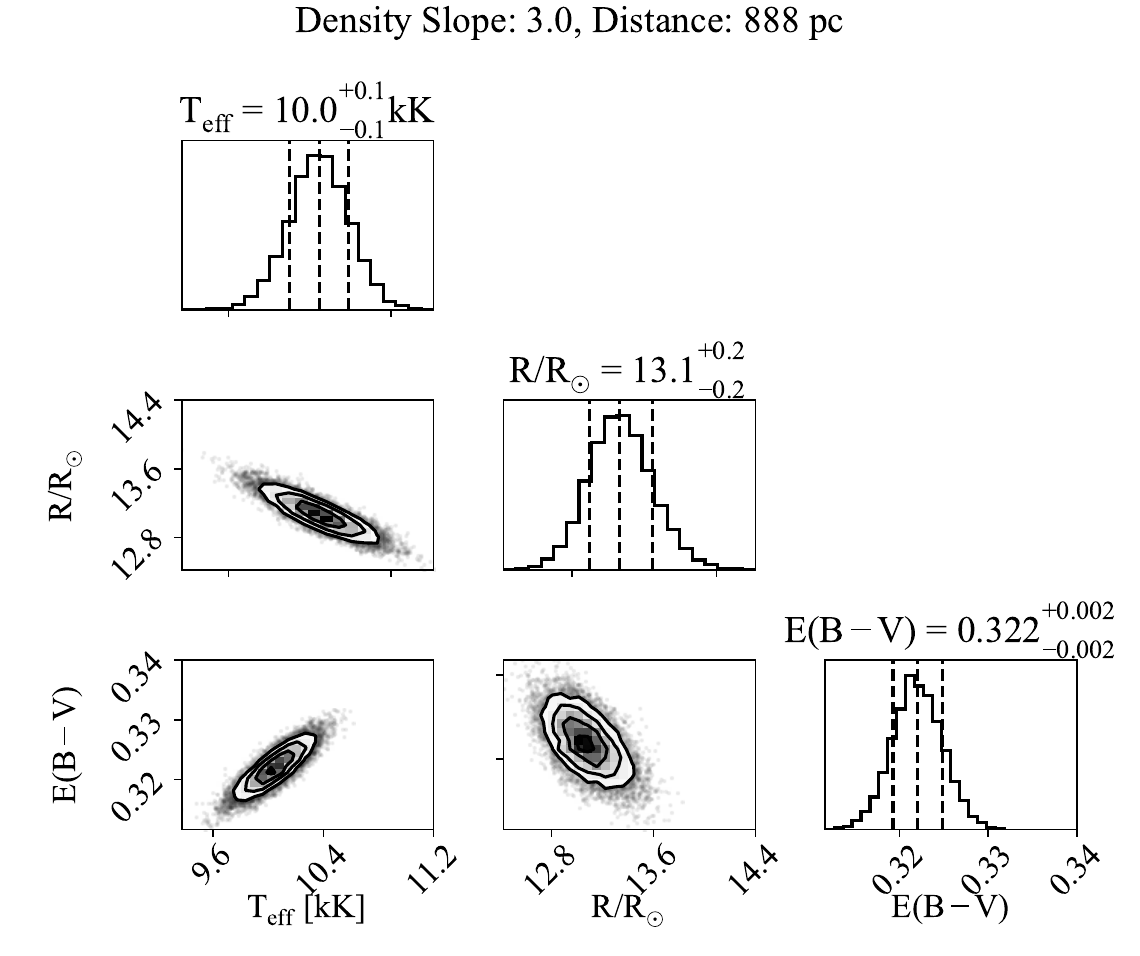}
\includegraphics[width=0.45\textwidth]{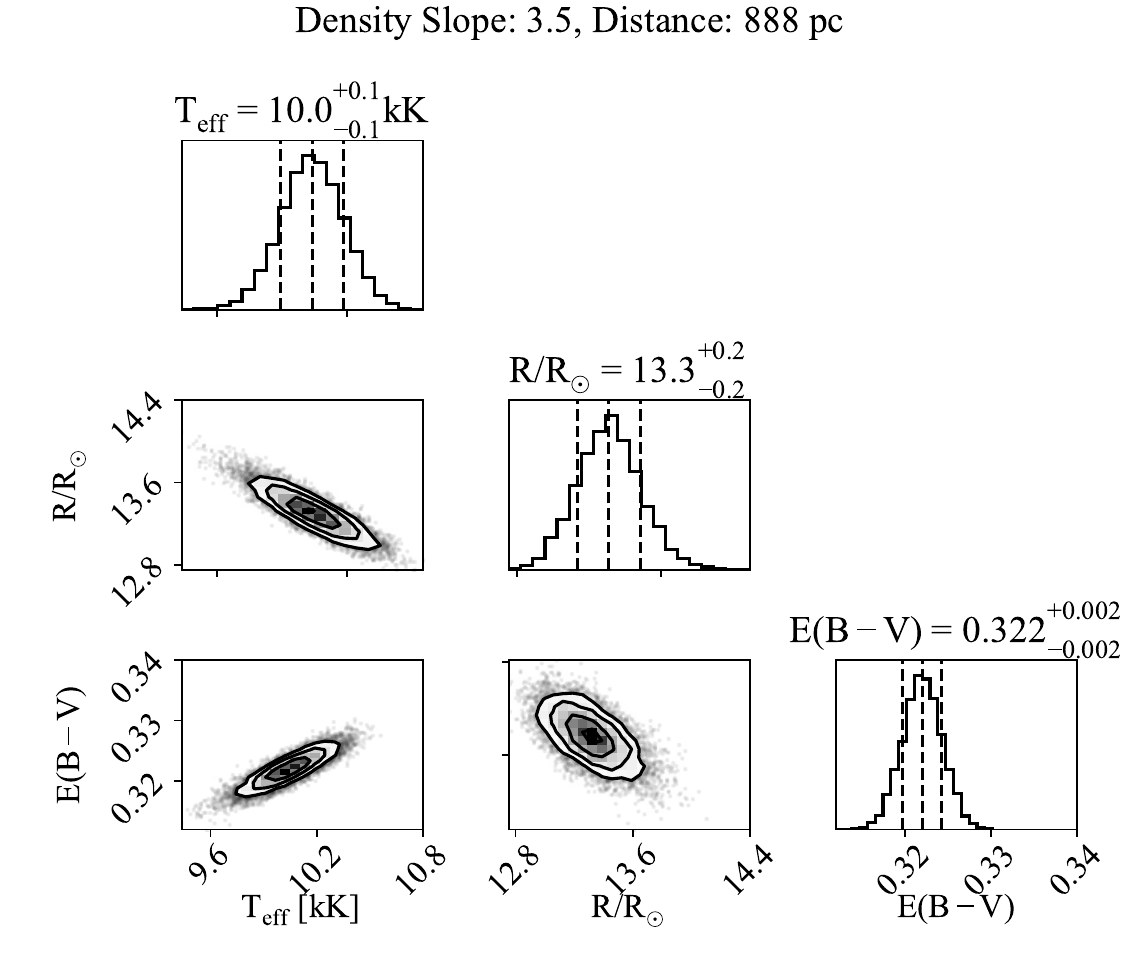}
\includegraphics[width=0.45\textwidth]{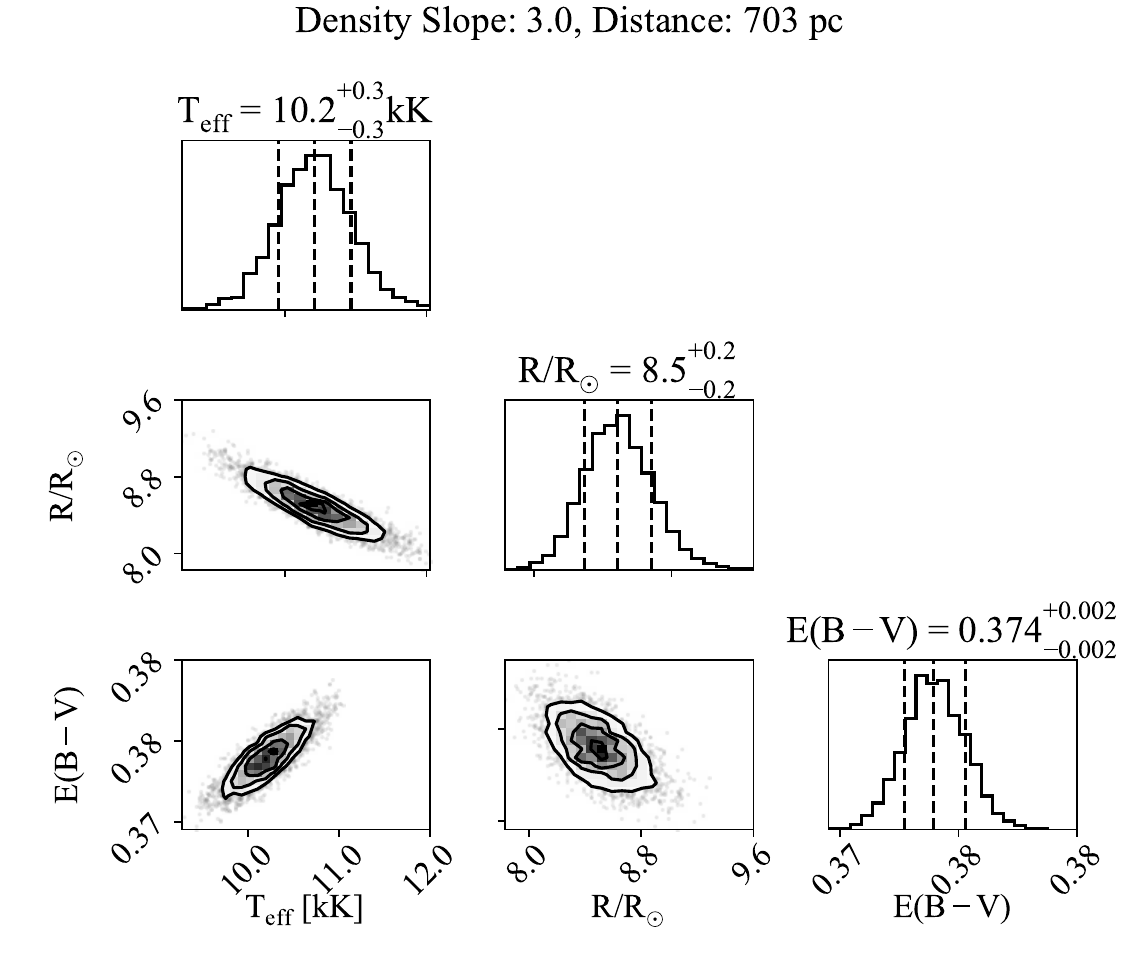}
\includegraphics[width=0.45\textwidth]{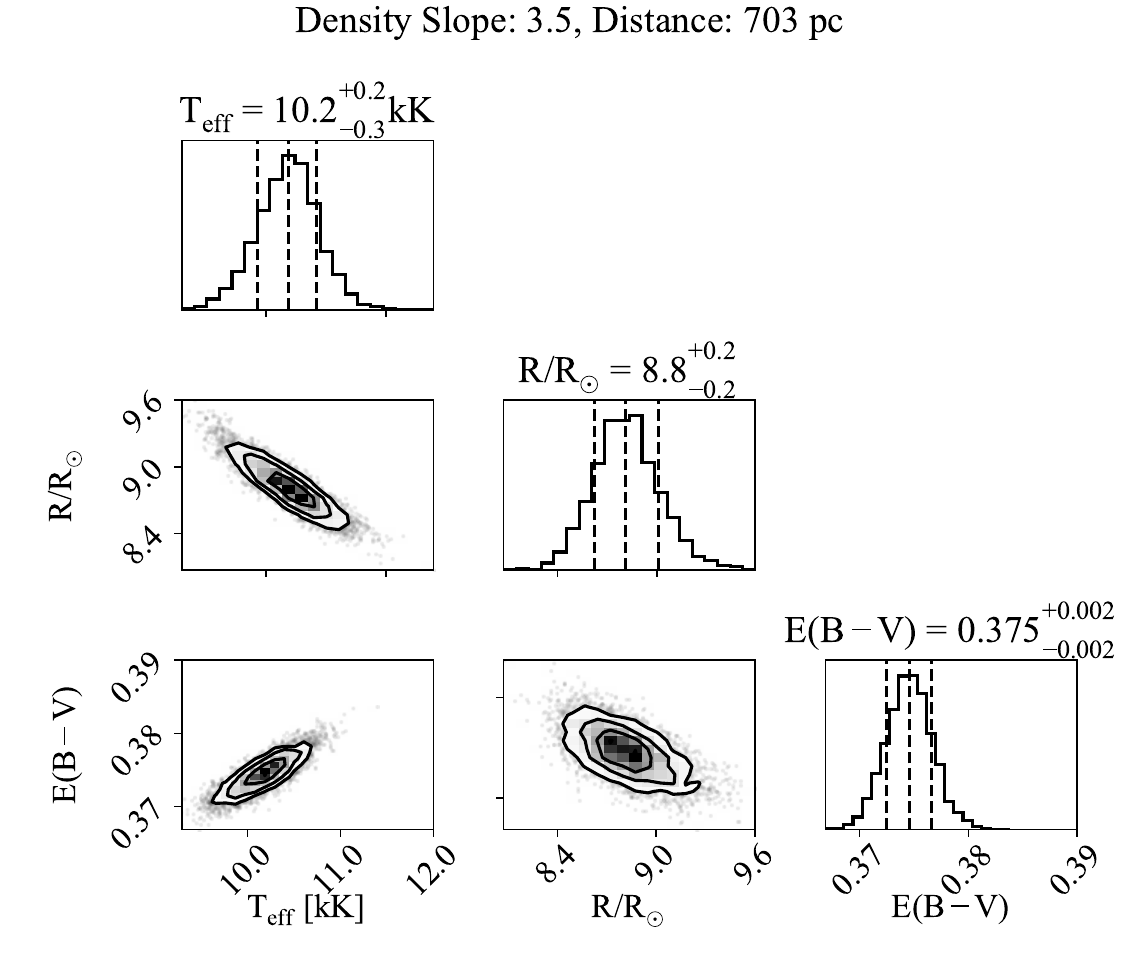}
\caption{Corner plots of the MCMC posteriors for the model with disk base density $\rho_0 = 5.0\times 10^{-12}\,{\rm g\,cm^{-3}}$.}
\label{fig:A2}
\end{figure*}

\begin{figure*}
\centering
\includegraphics[width=0.45\textwidth]{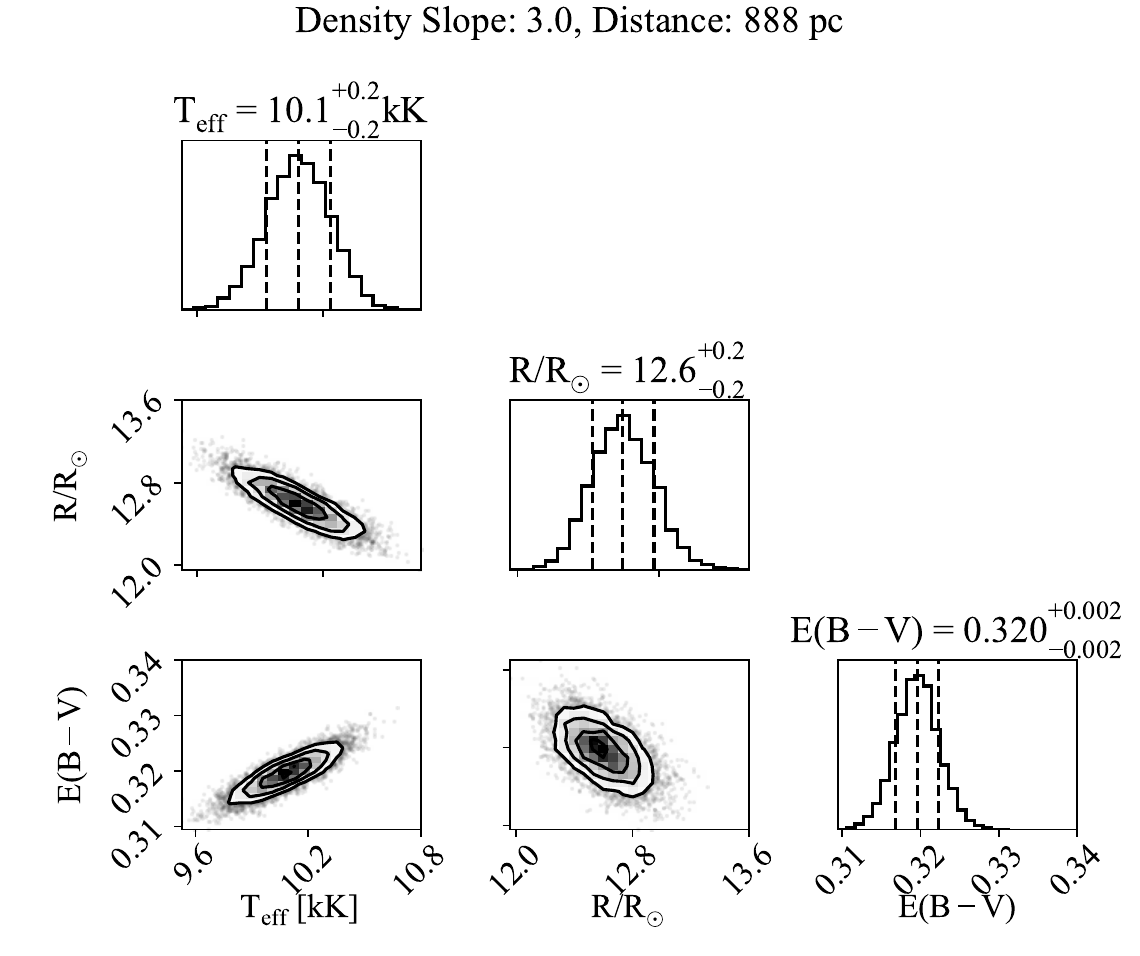}
\includegraphics[width=0.45\textwidth]{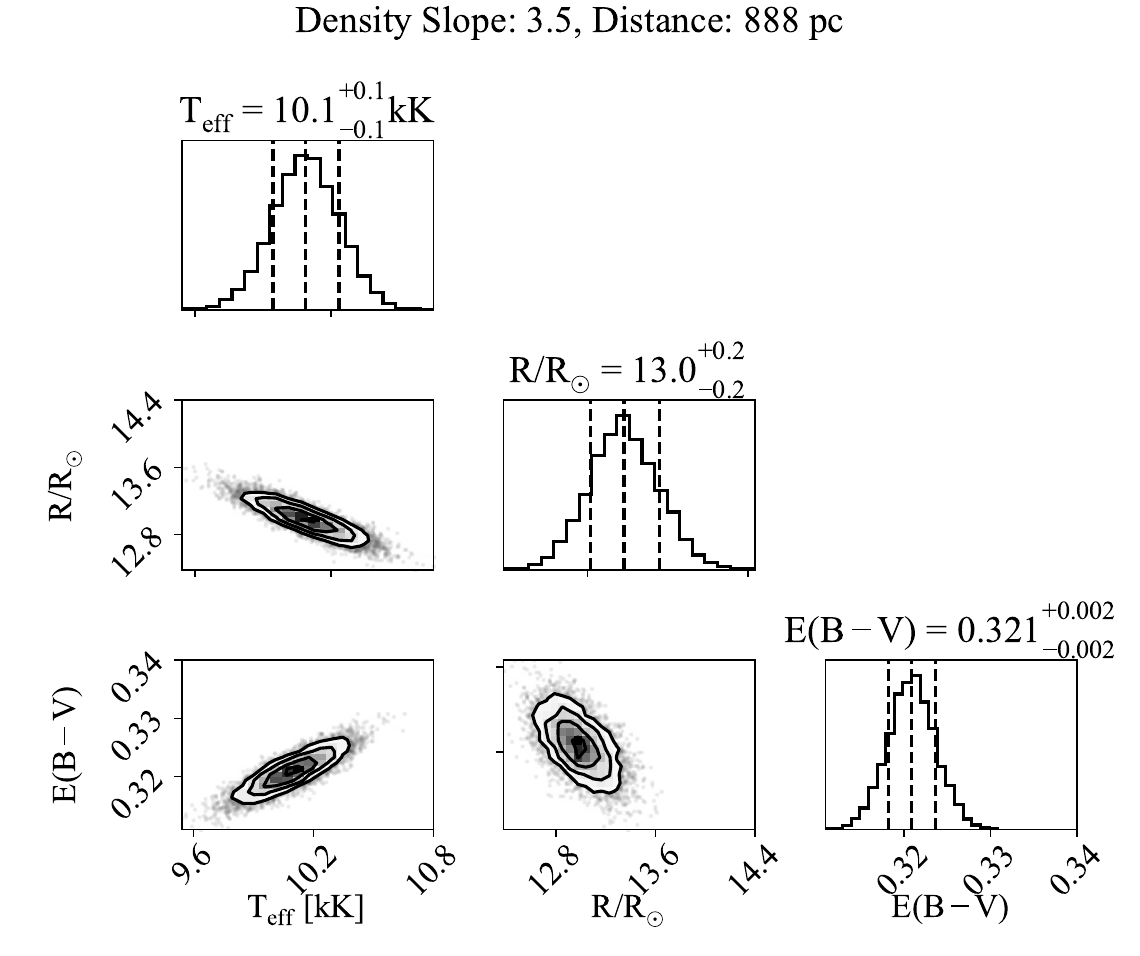}
\includegraphics[width=0.45\textwidth]{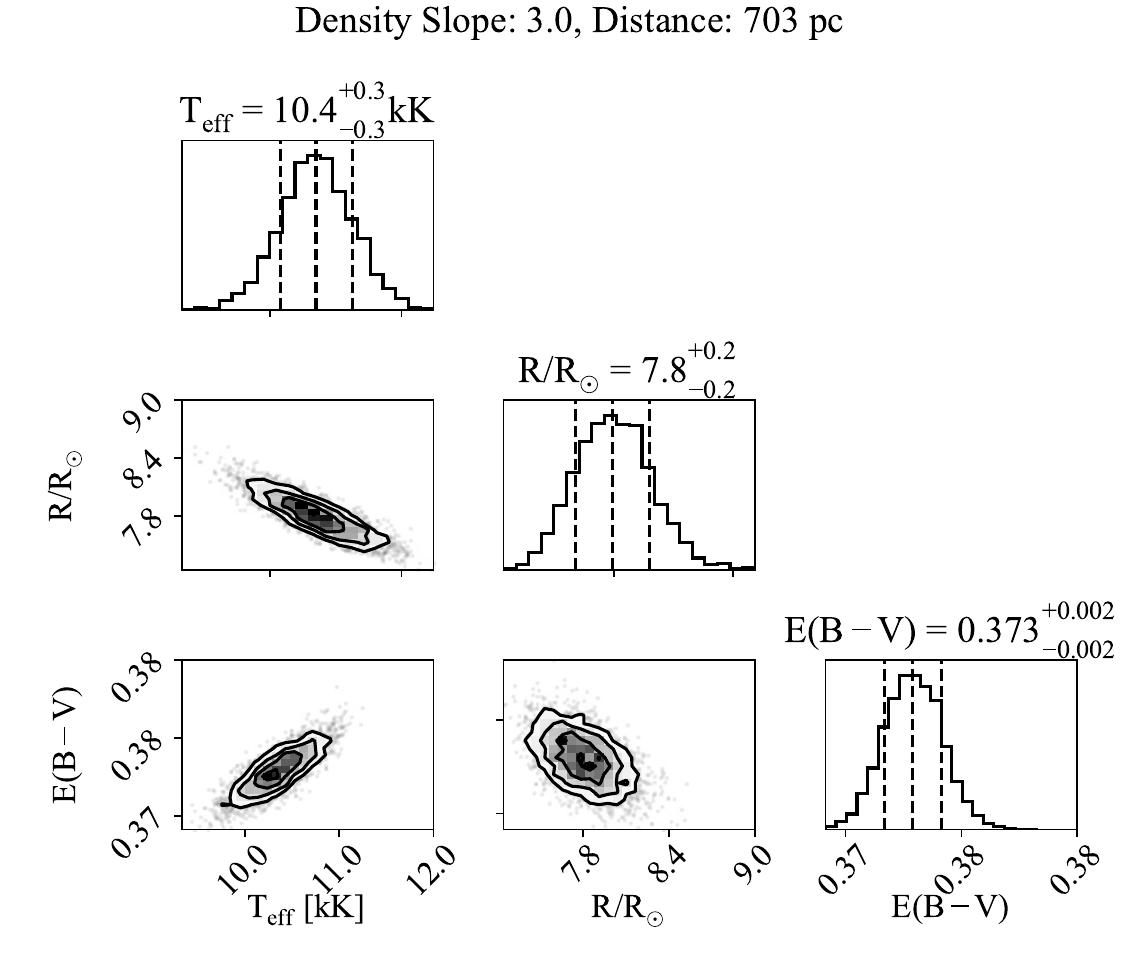}
\includegraphics[width=0.45\textwidth]{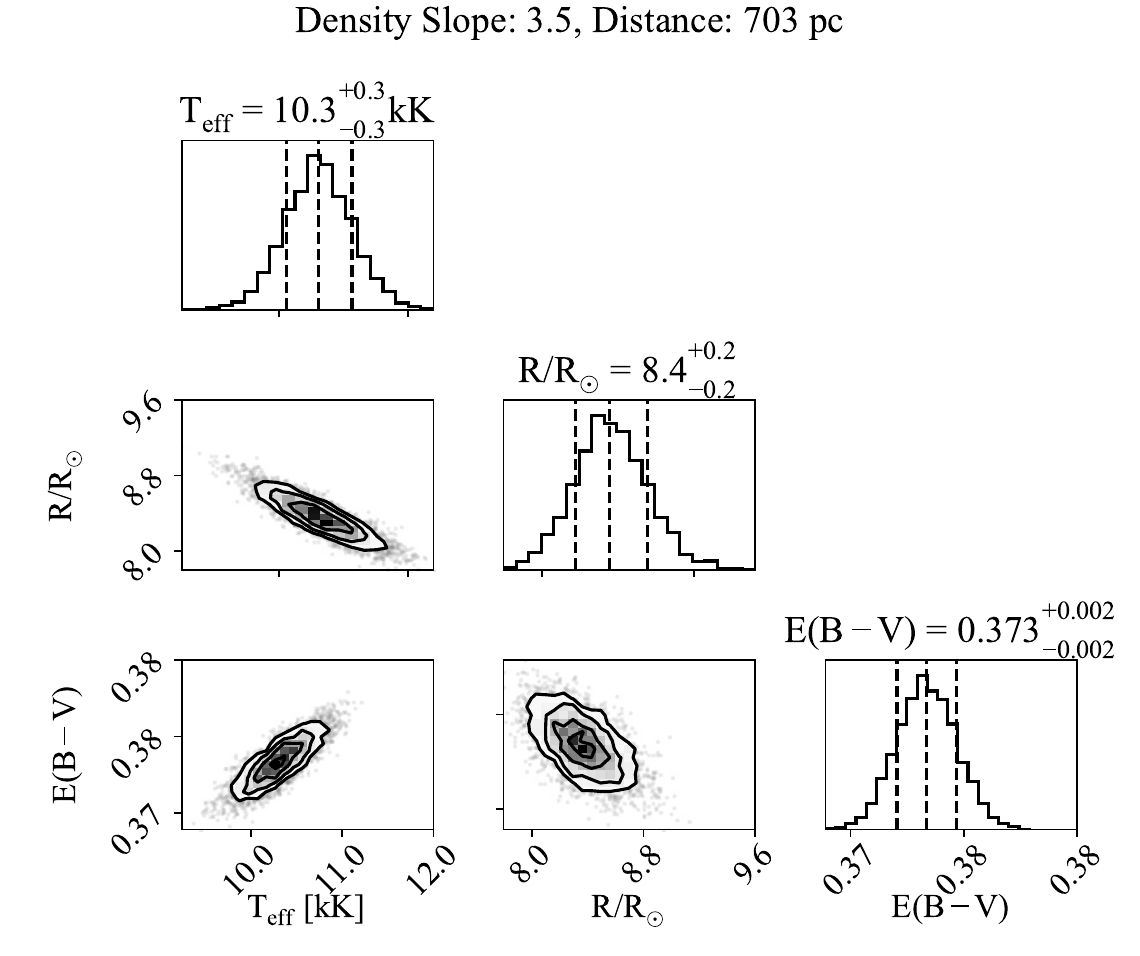}
\caption{Corner plots of the MCMC posteriors for the model with disk base density $\rho_0 = 7.5 \times 10^{-12}\,{\rm g\,cm^{-3}}$.}
\label{fig:A3}
\end{figure*}

\begin{figure*}
\centering
\includegraphics[width=0.45\textwidth]{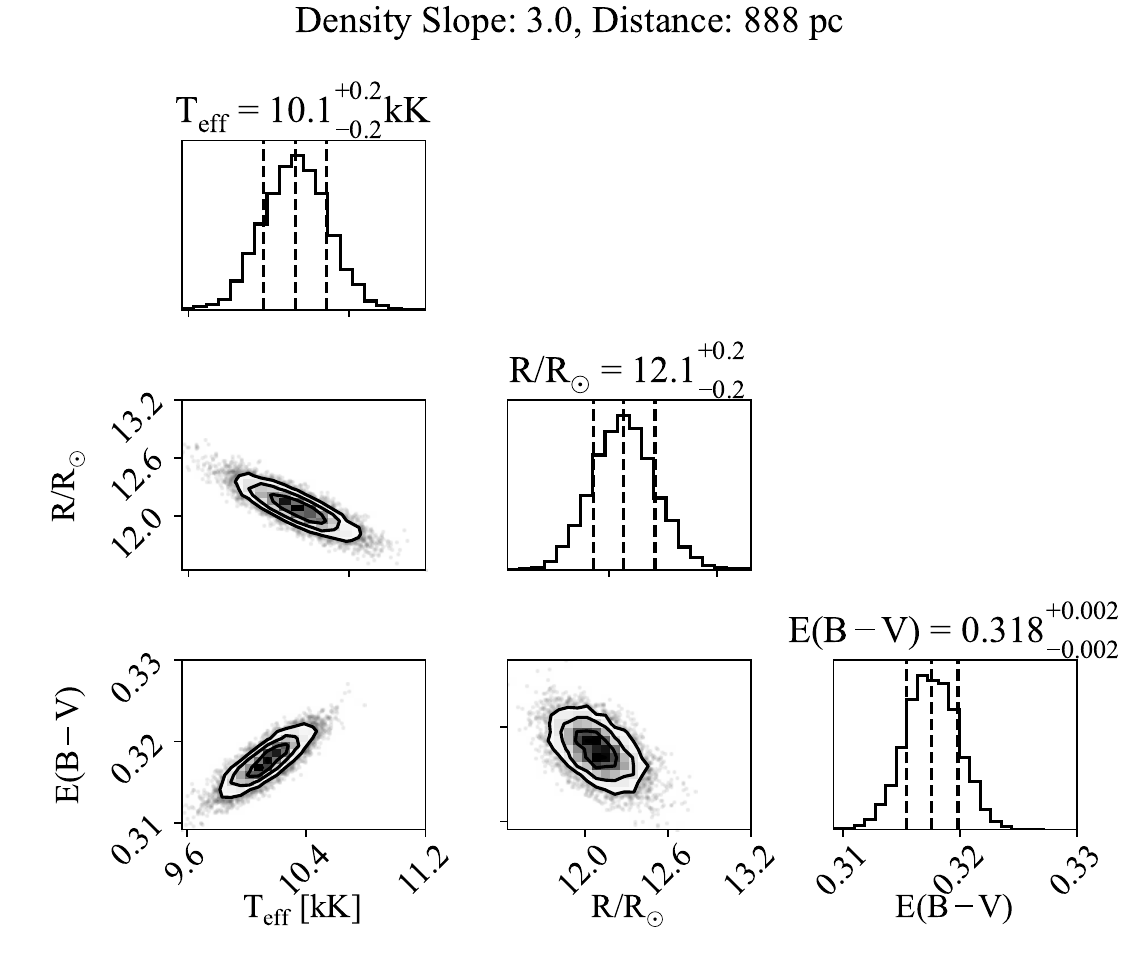}
\includegraphics[width=0.45\textwidth]{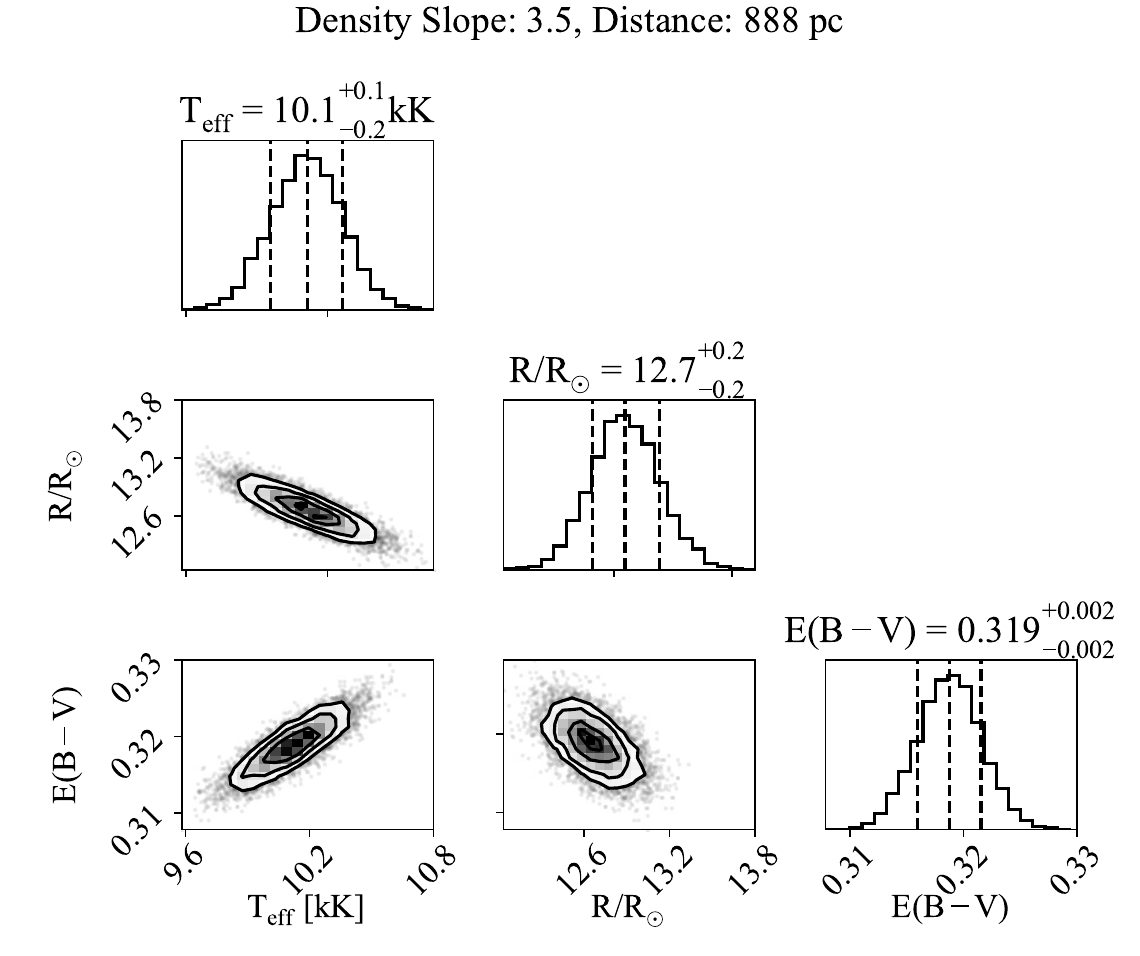}
\includegraphics[width=0.45\textwidth]{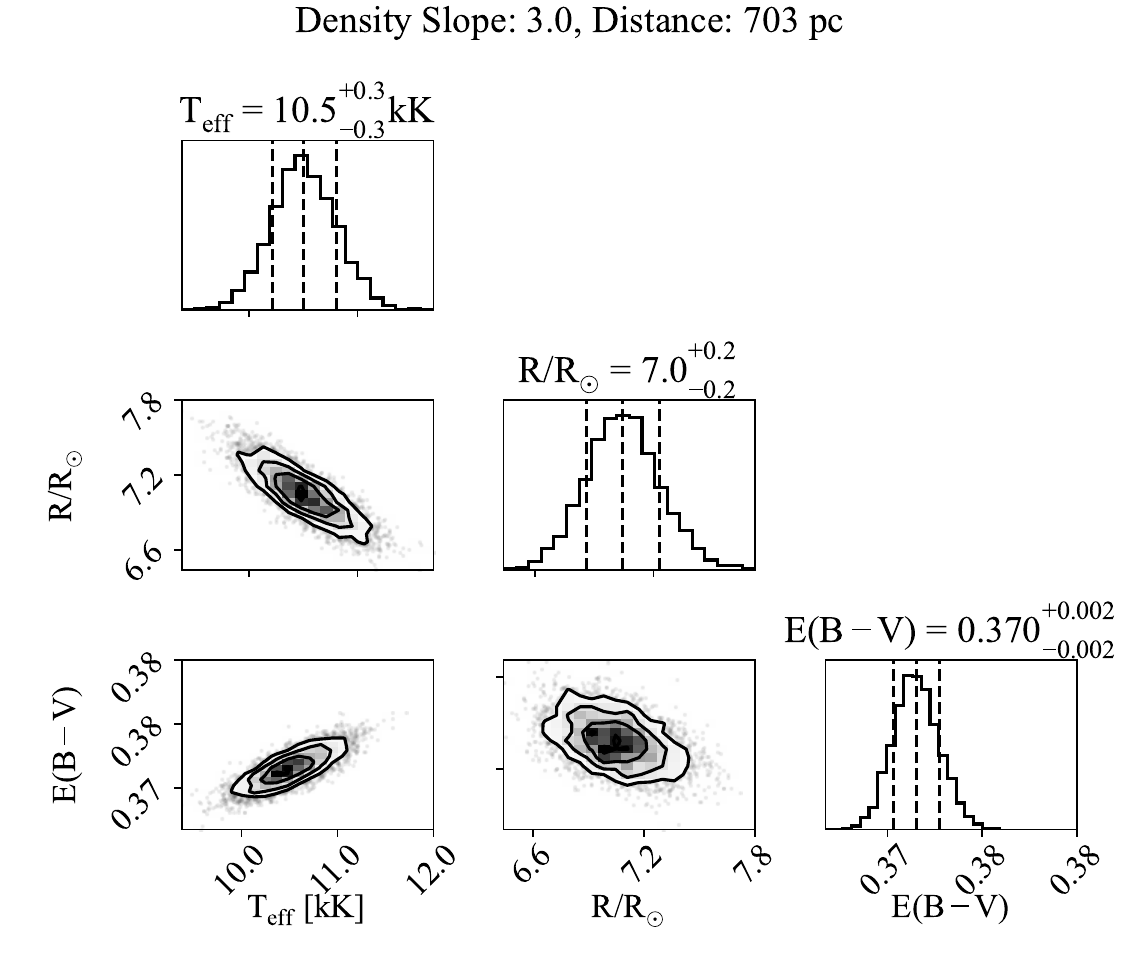}
\includegraphics[width=0.45\textwidth]{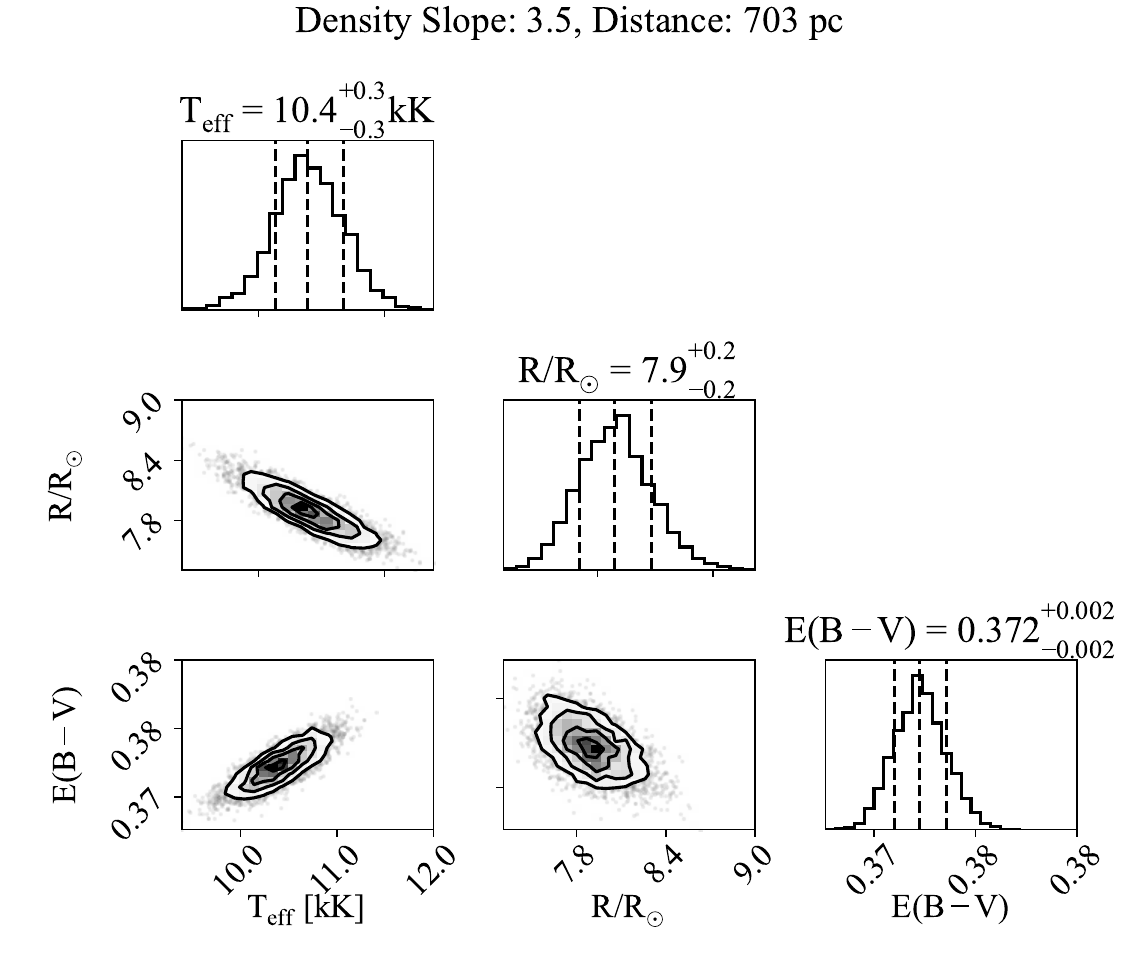}
\caption{Corner plots of the MCMC posteriors for the model with disk base density $\rho_0 = 10.0 \times 10^{-12}\,{\rm g\,cm^{-3}}$.}
\label{fig:A4}
\end{figure*}

\begin{figure*}
\centering
\includegraphics[width=0.45\textwidth]{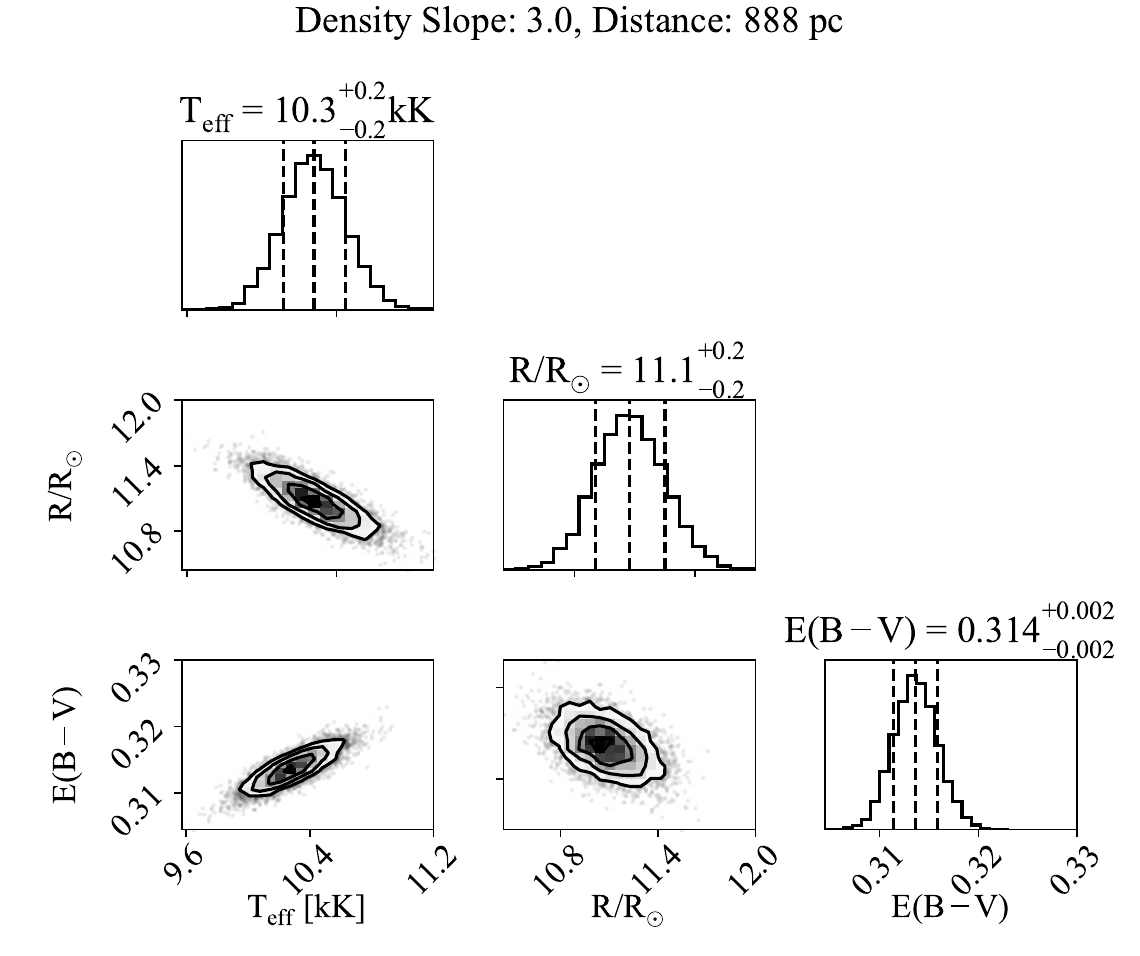}
\includegraphics[width=0.45\textwidth]{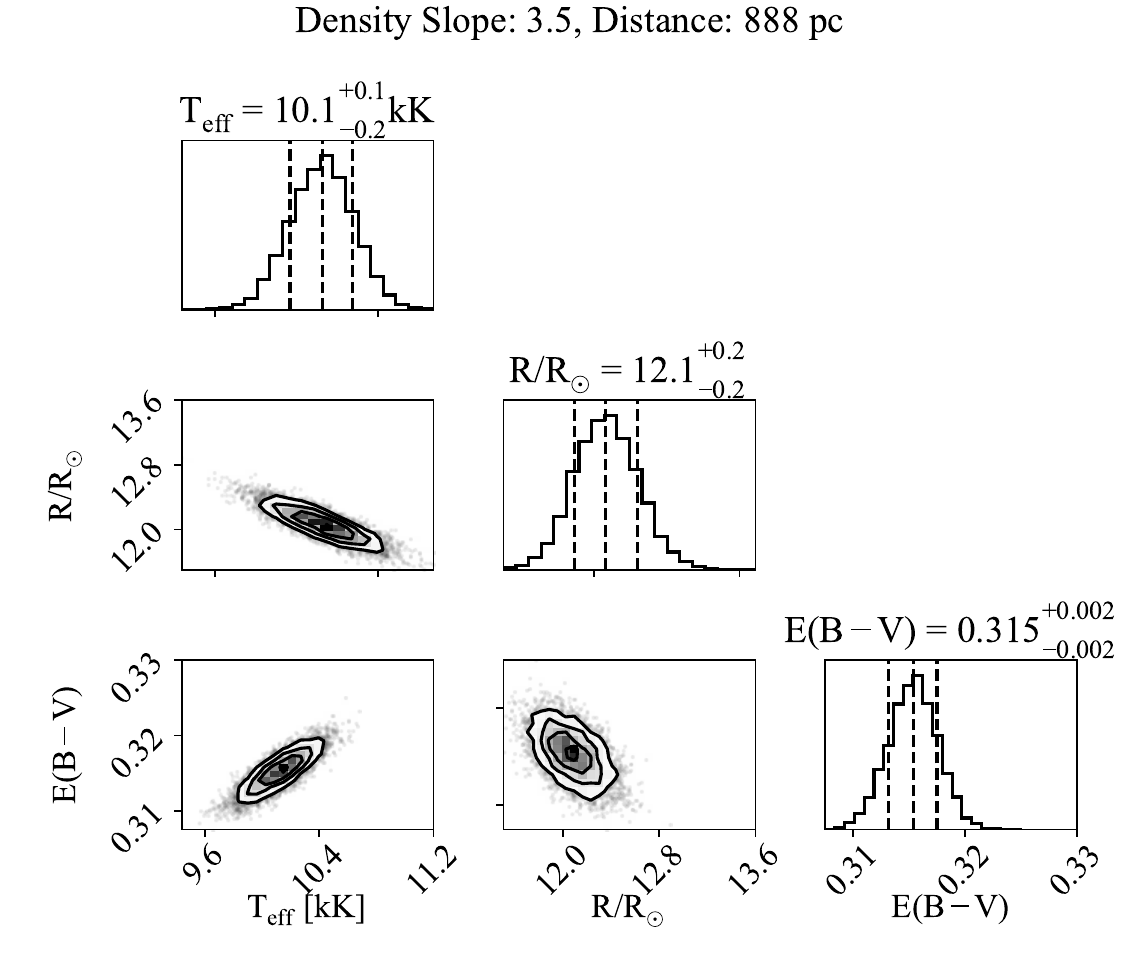}
\includegraphics[width=0.45\textwidth]{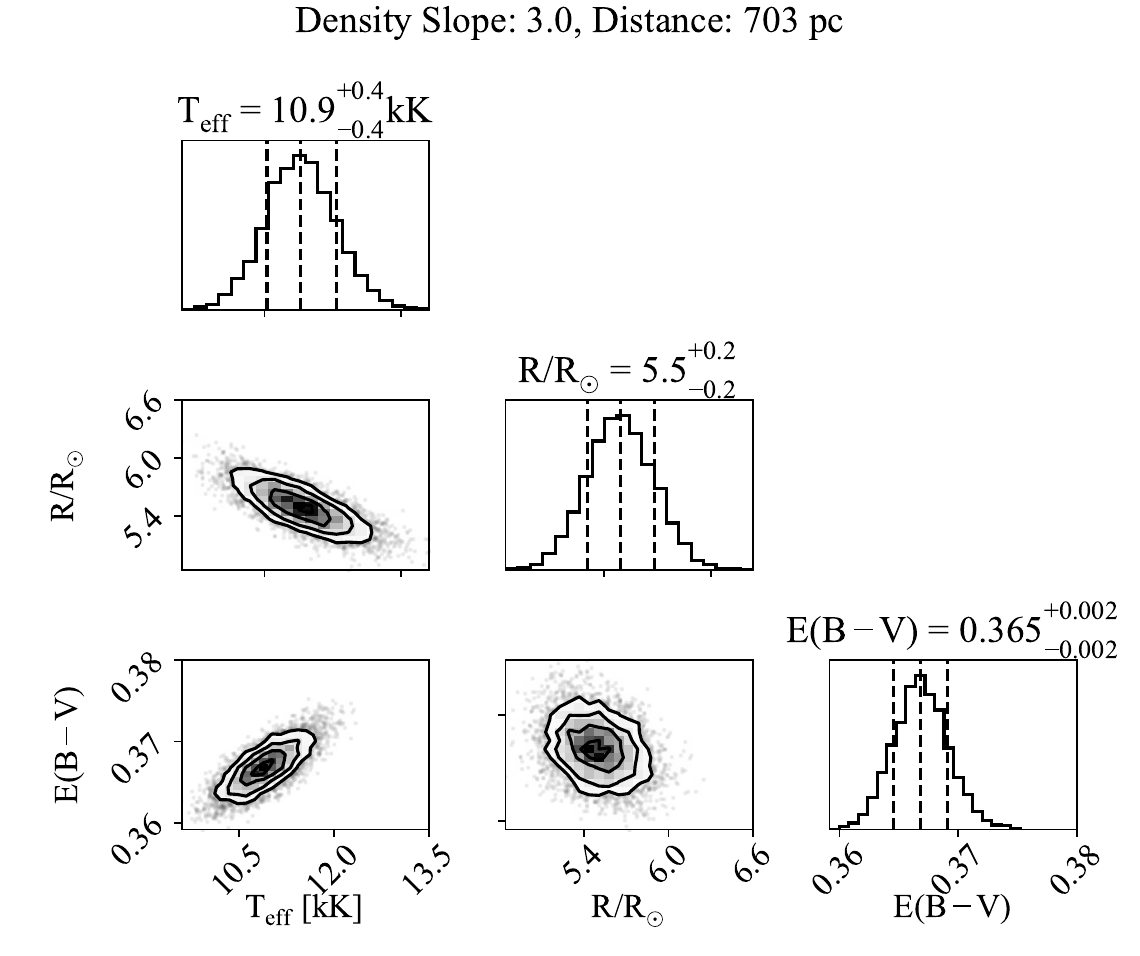}
\includegraphics[width=0.45\textwidth]{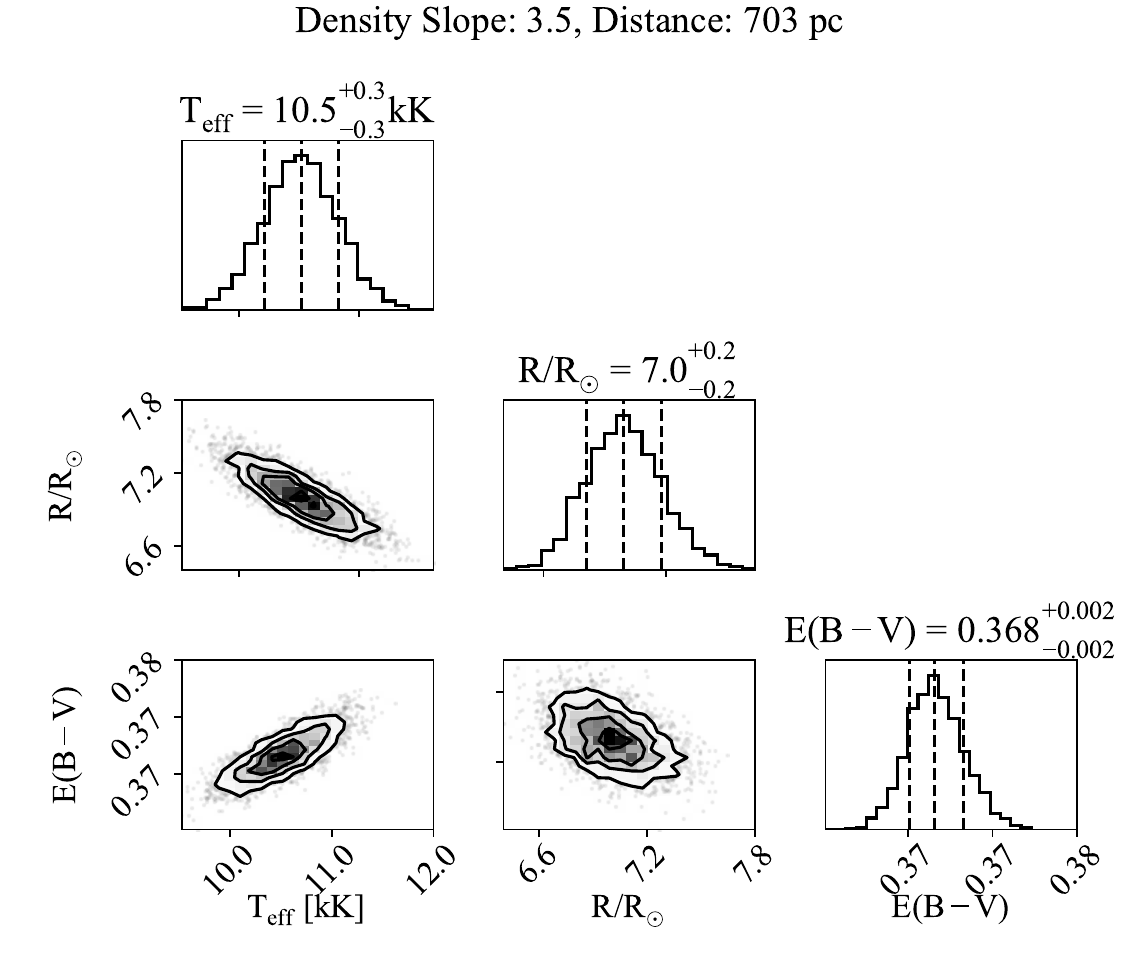}
\caption{Corner plots of the MCMC posteriors for the model with disk base density $\rho_0 = 15.0 \times 10^{-12}\,{\rm g\,cm^{-3}}$.}
\label{fig:A5}
\end{figure*}

\begin{figure*}
\centering
\includegraphics[width=0.45\textwidth]{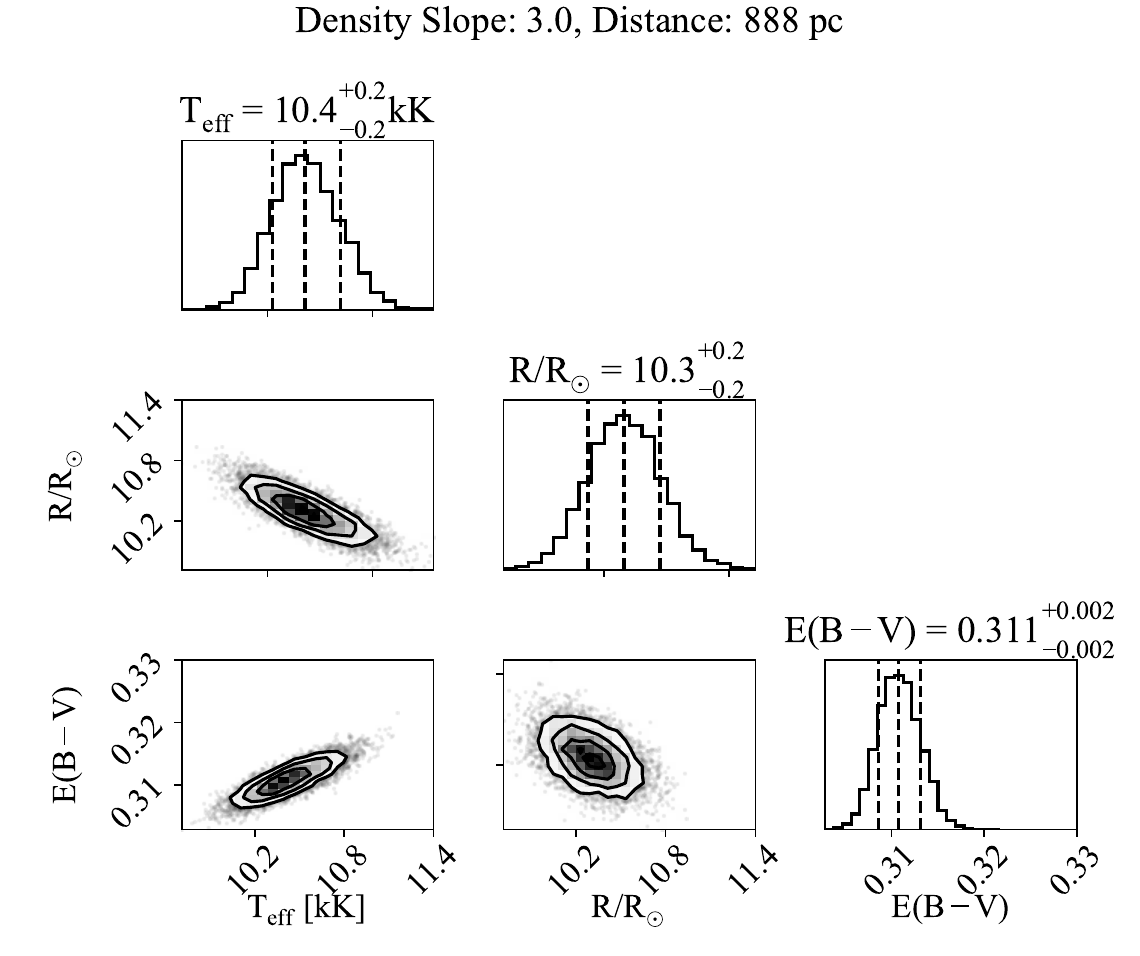}
\includegraphics[width=0.45\textwidth]{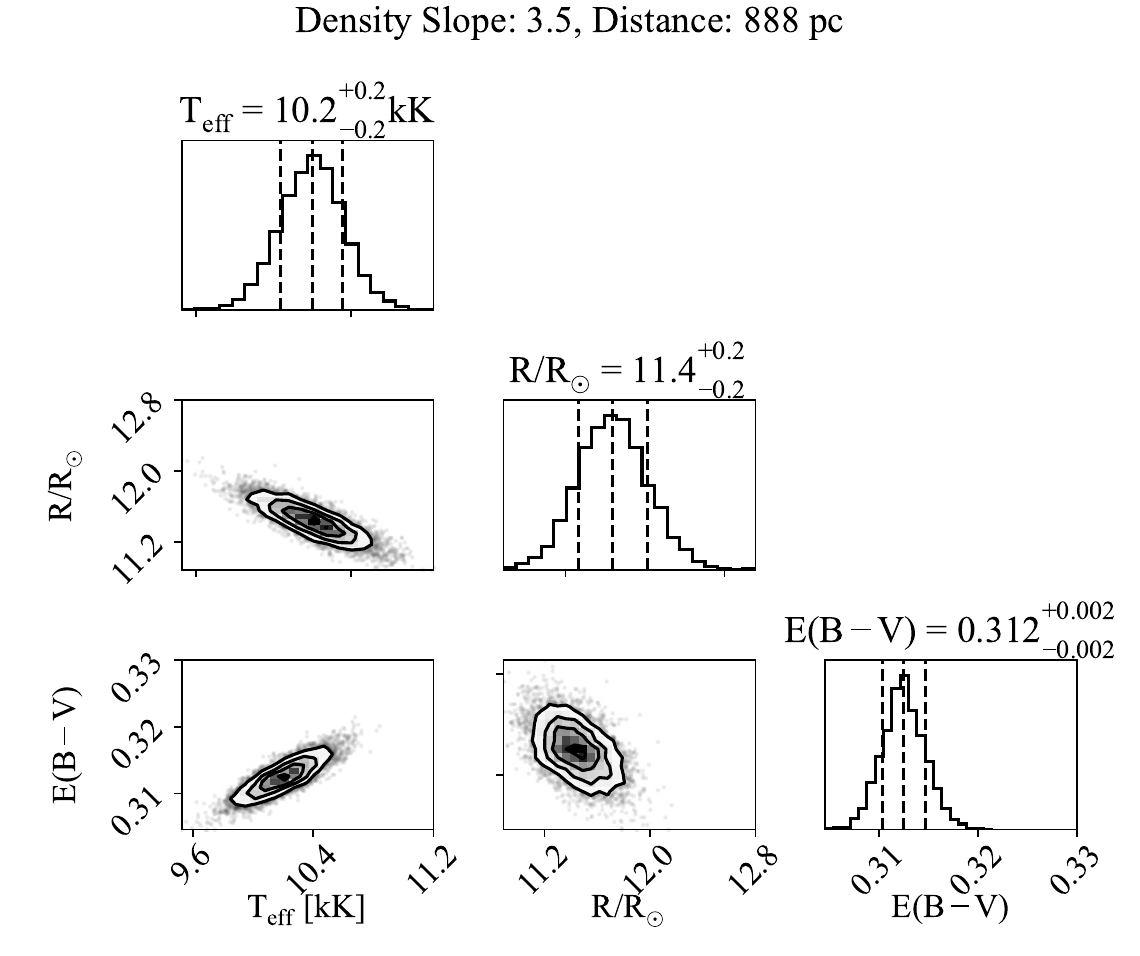}
\includegraphics[width=0.45\textwidth]{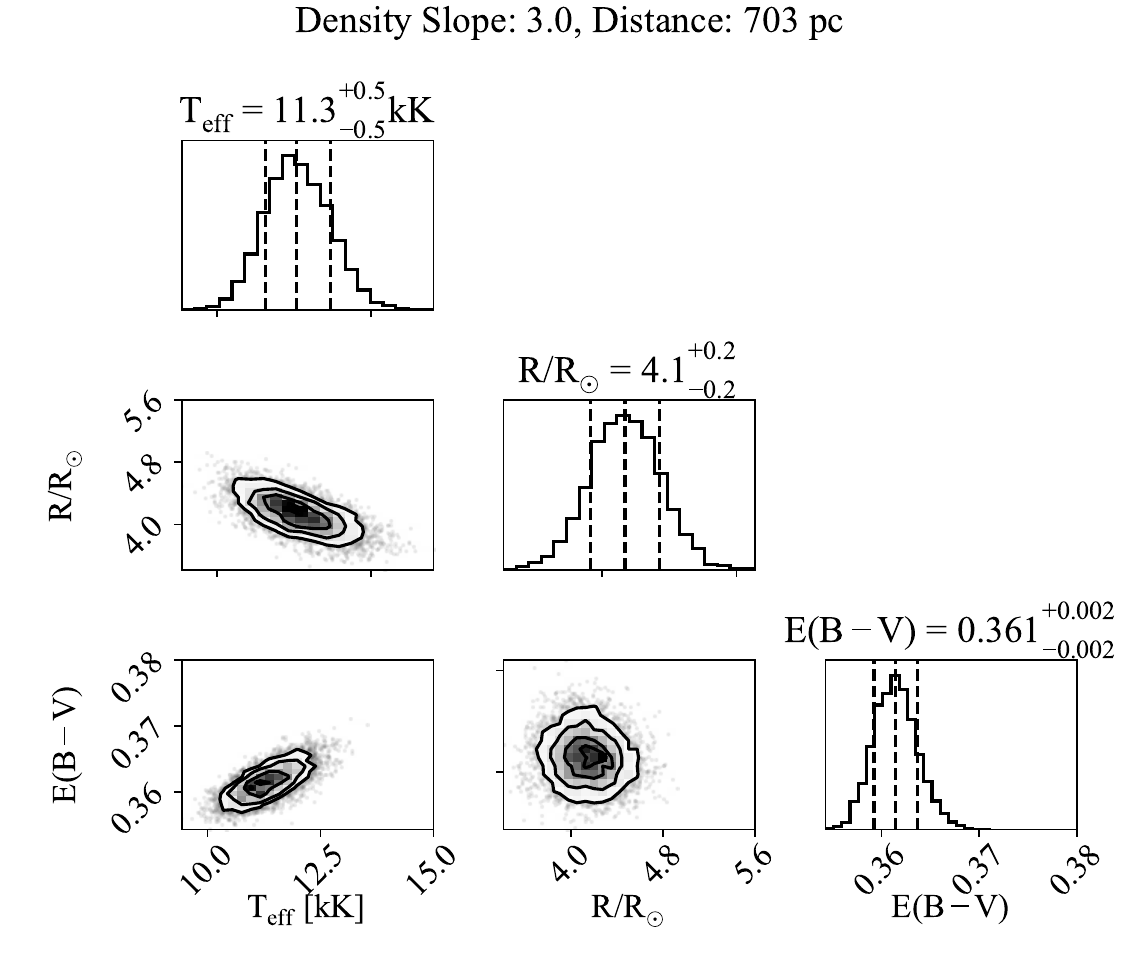}
\includegraphics[width=0.45\textwidth]{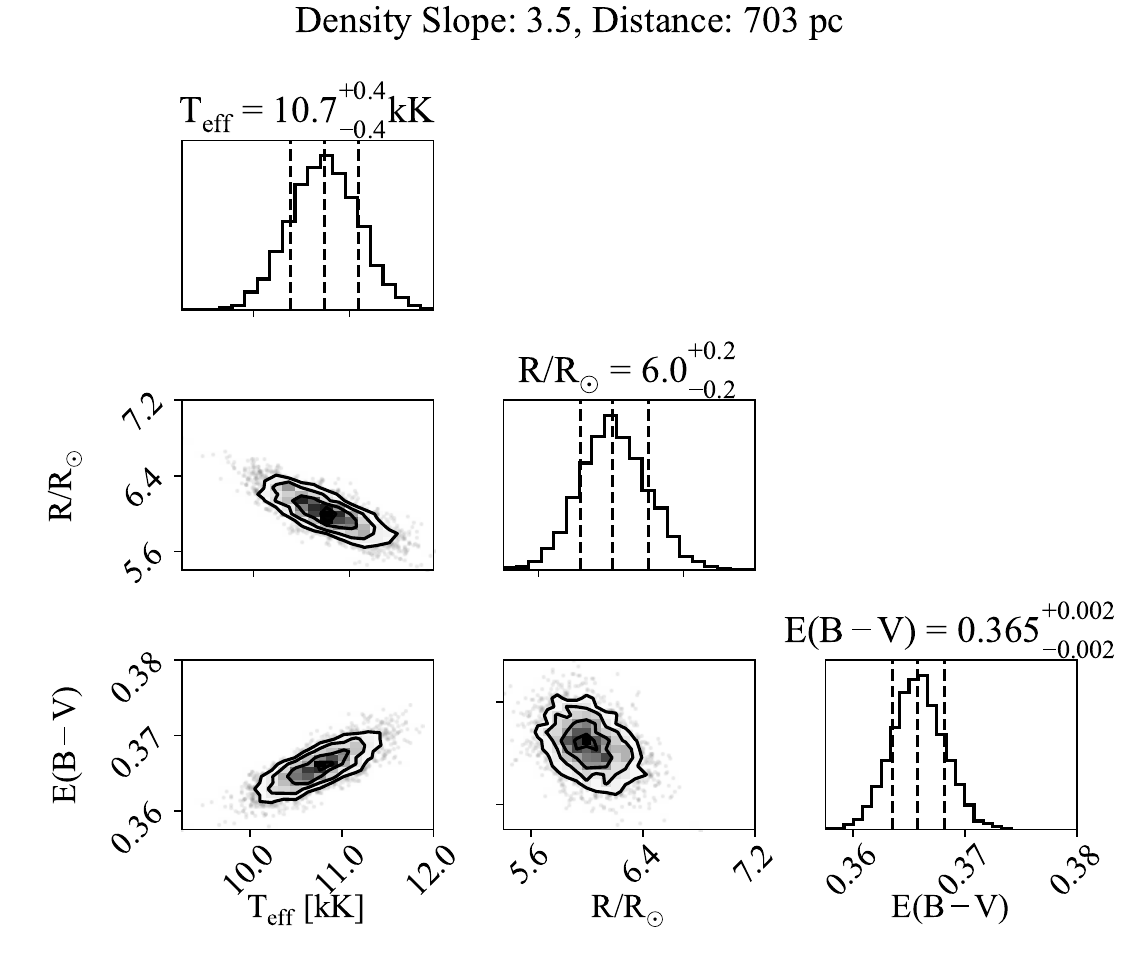}
\caption{Corner plots of the MCMC posteriors for the model with disk base density $\rho_0 = 20.0 \times 10^{-12}\,{\rm g\,cm^{-3}}$.}
\label{fig:A6}
\end{figure*}

\begin{figure*}
\centering
\includegraphics[width=0.45\textwidth]{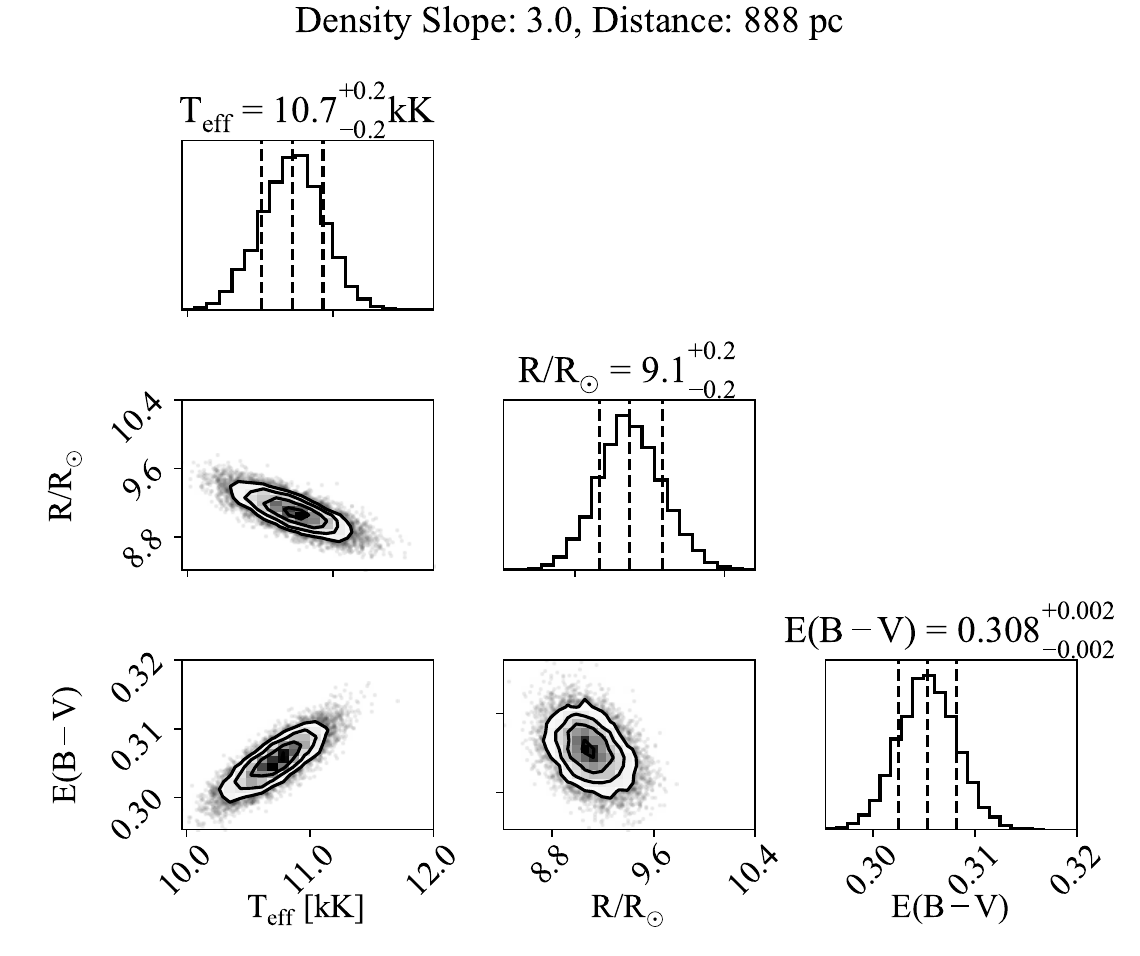}
\includegraphics[width=0.45\textwidth]{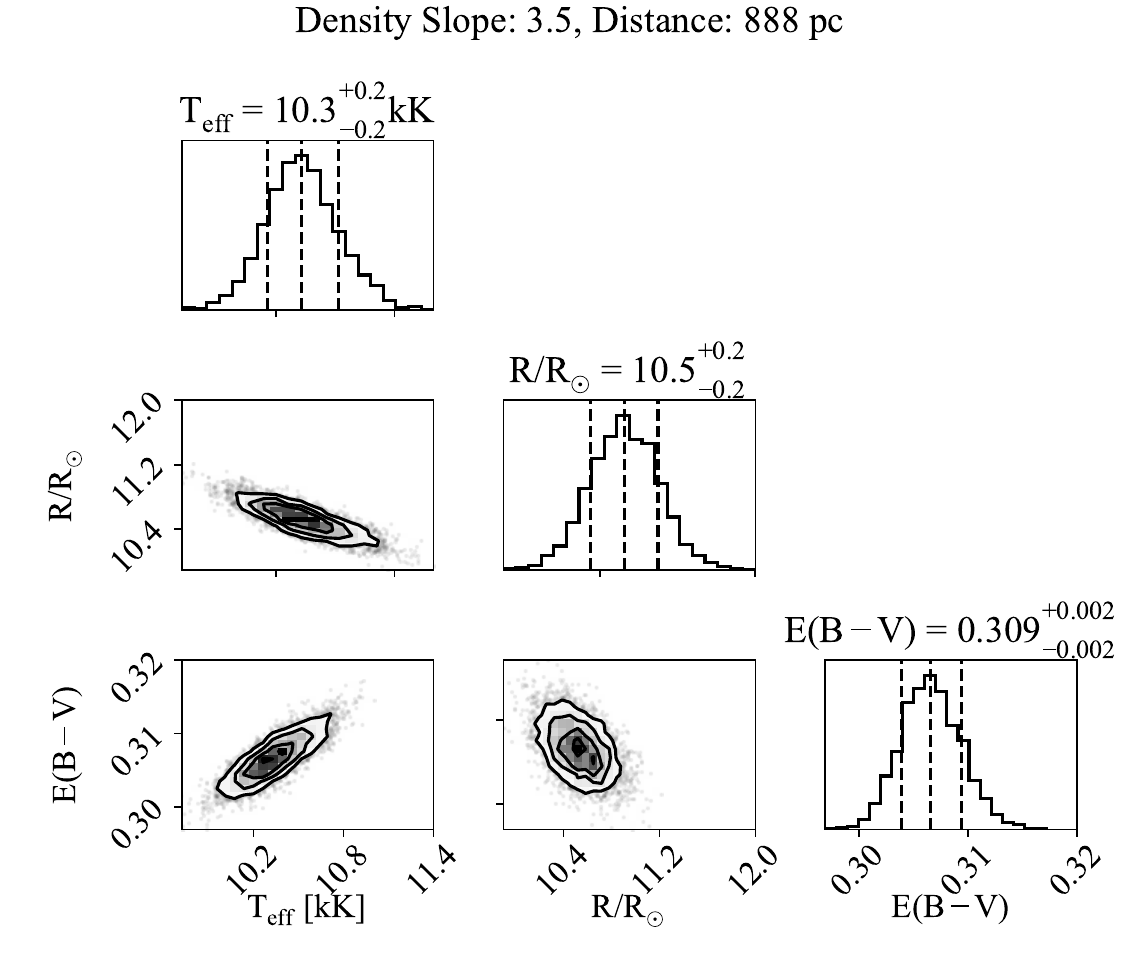}
\includegraphics[width=0.45\textwidth]{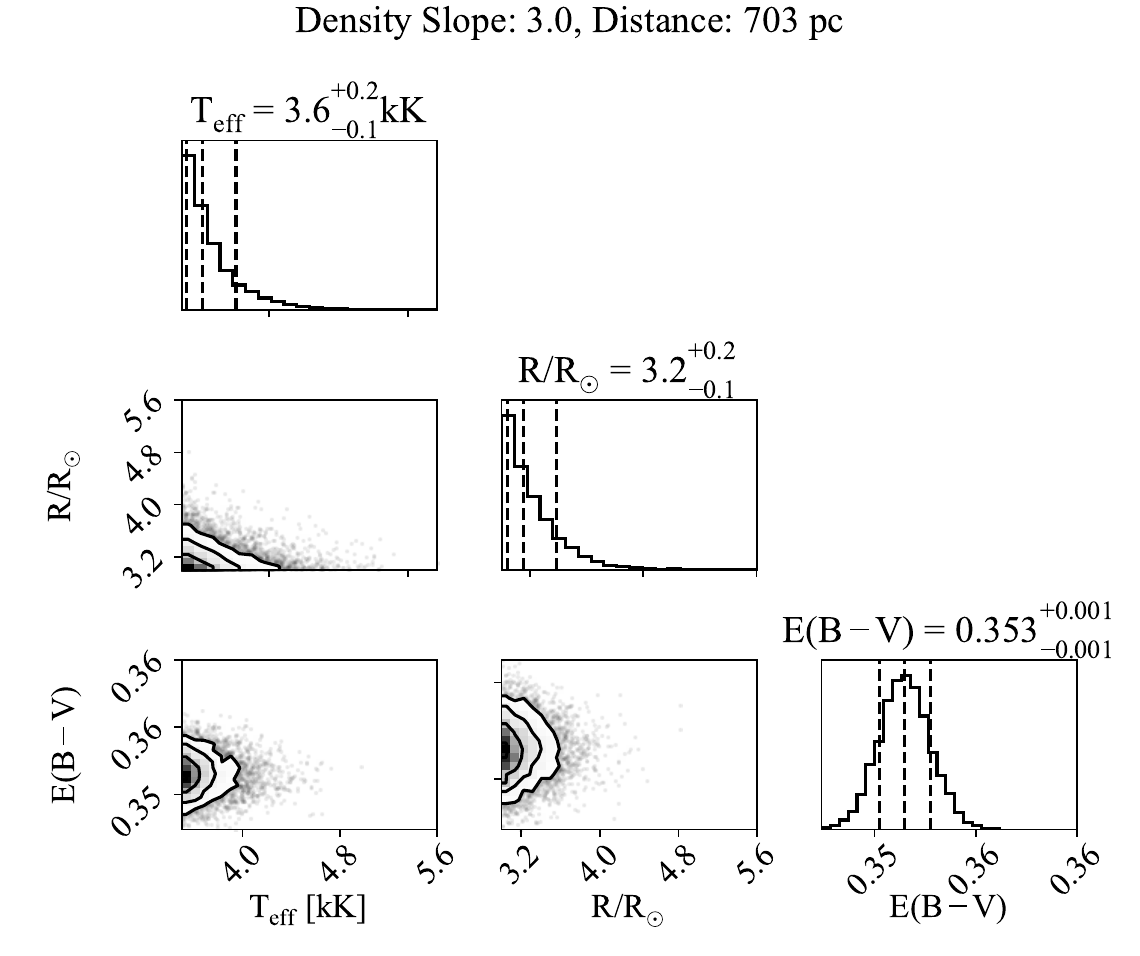}
\includegraphics[width=0.45\textwidth]{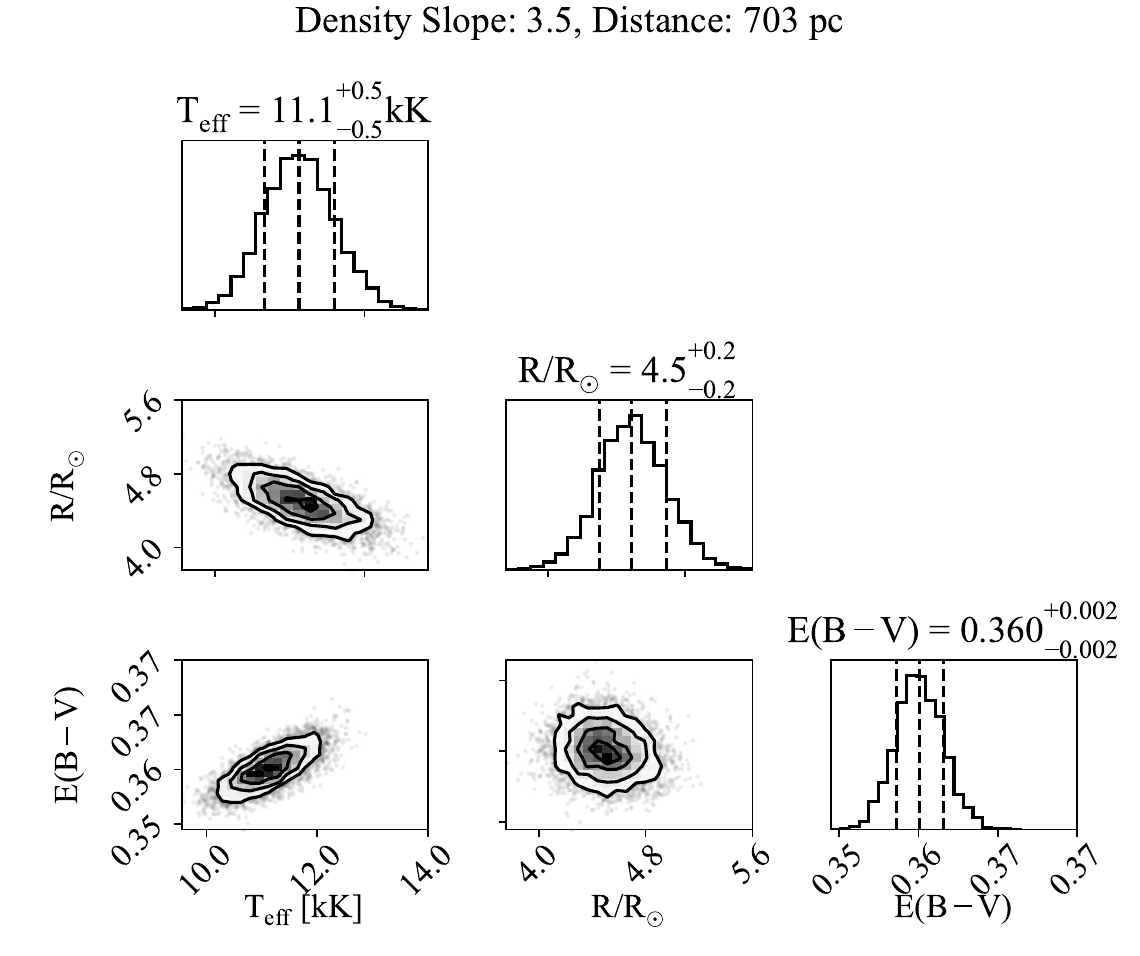}
\caption{Corner plots of the MCMC posteriors for the model with disk base density $\rho_0 = 30.0 \times 10^{-12}\,{\rm g\,cm^{-3}}$.}
\label{fig:A7}
\end{figure*}

Corner plots show the posterior probability distributions and parameter correlations from our MCMC analysis. Each diagonal panel displays the marginalized posterior distribution for a single parameter as a histogram, with the median and 68\% confidence intervals (16th to 84th percentiles) displayed in the panel titles above each histogram. Off-diagonal panels show the two-dimensional posterior distributions as contour plots. The contours represent the 16th, 50th (median), and 84th percentiles of the joint posterior distribution, corresponding to approximately 1$\sigma$ and 2$\sigma$ confidence regions for a Gaussian distribution. Each of the Figs.~\ref{fig:A1}–\ref{fig:A7} contains four panels for a single value of disk base density. Panels are arranged by distance $d=888$~pc (top row) and $d=703$~pc (bottom row) and by disk density slope $n=3.0$ (left column) and $n=3.5$ (right column).

\bibliography{references}{}
\bibliographystyle{aasjournal}



\end{document}